\documentclass[12pt,a4paper]{article}
\usepackage[utf8]{inputenc}
\usepackage{scalerel,stackengine,amsmath}
\usepackage{amsfonts}
\usepackage{amssymb}
\usepackage{hyphenat}
\usepackage{enumitem}
\usepackage{dsfont}
\usepackage[left=2cm,right=2cm,top=2.5cm,bottom=2.5cm]{geometry}
\usepackage{cite}
\usepackage{slashed}
\usepackage{hyperref}
\usepackage{graphicx}
\usepackage{caption}
\usepackage{float}
\usepackage{subcaption}
\usepackage{soul}
\usepackage[labelformat=simple]{subcaption}

\usepackage{authblk}
\hypersetup{
	unicode=true,          
	pdftoolbar=true,        
	pdfmenubar=true,        
	pdffitwindow=false,     
	pdfstartview={FitH},    
	pdftitle={DoFsLowScaleLepto},    
	pdfauthor={},     
	pdfsubject={},   
	pdfcreator={},   
	pdfproducer={Producer}, 
    pdfkeywords={Resonant Leptogenesis} {Baryon Asymmetry} {Transport Equations} {Degrees of Freedom},
    pdfnewwindow=true,      
	colorlinks=true,       
	linkcolor=blue,          
	citecolor=red,        
	filecolor=black,      
	urlcolor=blue           
}

\newcommand{\m}{{\bf m}}
\newcommand{\h}{{\bf h}}

\newcommand{\Id}{{\bf 1}}

\usepackage{ifthen}
\usepackage{tikz}
\usepackage{xspace}
\usepackage{physics,bm}

\newcommand{\GeV}{{\rm GeV}\xspace}
\newcommand{\TeV}{{\rm TeV}\xspace}

\newcommand{\bvev}[1]{\Big\langle #1 \Big\rangle}

\newcommand{\lrb}[1]{\left( #1 \right)}
\newcommand{\lrsb}[1]{\left[ #1 \right]}
\newcommand{\lrcb}[1]{\left\{ #1 \right\}}

\renewcommand{\Re}[1]{\operatorname{Re}\lrsb{#1}\xspace}
\renewcommand{\Im}[1]{\operatorname{Im}\lrsb{#1}\xspace}

\renewcommand{\Tr}[1]{\operatorname{Tr}\lrsb{#1}\xspace}

\newcounter{NumArgs}

\newcommand{\eqs}[1]{\setcounter{NumArgs}{0}\foreach\i in{#1}{\stepcounter{NumArgs}}%
	\ifthenelse{\equal{\theNumArgs}{1}}{(\ref{#1})}%
	{\ifthenelse{\equal{\theNumArgs}{2}}%
		{\foreach\i[count=\q]in{#1}{\ifthenelse{\equal{\q}{\theNumArgs}}{and (\ref{\i})}{(\ref{\i})~}}}%
		{\foreach\i[count=\q]in{#1}{\ifthenelse{\equal{\q}{\theNumArgs}}{and (\ref{\i})}{(\ref{\i}),~}}}}}

\newcommand{\refs}[1]{\setcounter{NumArgs}{0}\foreach\i in{#1}{\stepcounter{NumArgs}}%
	\ifthenelse{\equal{\theNumArgs}{1}}{(\ref{#1})}%
	{\ifthenelse{\equal{\theNumArgs}{2}}%
		{\foreach\i[count=\q]in{#1}{\ifthenelse{\equal{\q}{\theNumArgs}}{and (\ref{\i})}{(\ref{\i})~}}}%
		{\foreach\i[count=\q]in{#1}{\ifthenelse{\equal{\q}{\theNumArgs}}{and (\ref{\i})}{(\ref{\i}),~}}}}}

\newcommand{\Figs}[1]{\setcounter{NumArgs}{0}\foreach\i in{#1}{\stepcounter{NumArgs}}%
	\ifthenelse{\equal{\theNumArgs}{1}}{Figure~\ref{#1}}%
	{\ifthenelse{\equal{\theNumArgs}{2}}%
		{Figures~\foreach\i[count=\q]in{#1}{\ifthenelse{\equal{\q}{\theNumArgs}}{and \ref{\i}}{\ref{\i}~}}}%
		{Figures~\foreach\i[count=\q]in{#1}{\ifthenelse{\equal{\q}{\theNumArgs}}{and \ref{\i}}{\ref{\i},~}}}}}

 \renewcommand{\theequation}{\arabic{section}.\arabic{equation}}

\usepackage{indentfirst}

\definecolor{ao}{rgb}{0.0, 0.0, 1.0}
\newenvironment{bl}[1]{{\color{ao}  #1}}

\newcommand*{\email}[1]{%
	\footnotesize\href{mailto:#1}{\bl{#1}}
}

\title{\textbf{Varying Entropy Degrees of Freedom Effects\\ in Low-Scale Leptogenesis}}

\author[]{Dimitrios Karamitros${}^{\,a,b,c\,}$\footnote{\email{dimitrios.d.karamitros@jyu.fi}} }
\author[]{Thomas McKelvey${}^{\,a\,}$\footnote{\email{thomas.mckelvey@manchester.ac.uk}} } 
\author[]{Apostolos Pilaftsis${}^{\,a\,}$\footnote{\email{apostolos.pilaftsis@manchester.ac.uk}} }

\affil[]{
\normalsize\textit{\hspace{1.4cm}${}^a$Department
 of Physics and Astronomy, University of Manchester,\newline Manchester, M13 9PL, United Kingdom}
 
\textit{\hspace{1.4cm}${}^b$Helsinki Institute of Physics, University of Helsinki, Helsinki,\newline P.O. Box 64, FIN-00014, Finland} 
 
\textit{\hspace{1.4cm}${}^c$Department of Physics, University of Jyv\"askyl\"a, Jyv\"askyl\"a,\newline  P.O.Box 35 (YFL), FIN-40014, Finland  } }

\date{\empty}

\begin{document}
\setcounter{page}{1}

{\let\newpage\relax\maketitle}
\maketitle

\flushbottom
\vspace{-1cm}
\begin{abstract}
    
\noindent 
We analyse in detail the effect of varying entropy degrees of freedom on low-scale lepto\-genesis models. As an archetypal model, we consider the Tri-Resonant Leptogensis${}$~(TRL) scenario introduced recently by the authors, where the neutrino-Yukawa coupling matrix is dictated by an approximate $\mathbb{Z}_n$ discrete symmetry (with $n=3,6$). TRL models exhibit no preferred direction in the leptonic flavour space and have the remarkable feature that leptogenesis can successfully take place even if all light neutrinos are strictly massless up to one-loop order. Most interestingly, for TRL scenarios with heavy Majorana neutrinos lighter than~100~GeV, temperature varying degrees of freedom associated with the entropy of the plasma have a dramatic impact on the predictions\- of the Baryon Asymmetry in the Universe (BAU), and may depend on the freeze-out sphaleron temperature $T_{\rm sph}$. We find that this is a generic feature of most freeze-out low-scale leptogenesis models discussed in the literature. In the same context, we consider heavy-neutrino scenarios realising dynamics related to critical unstable qudits in the thermal plasma and assess their significance in generating the~BAU. The phenomenological implications of TRL scenarios at the intensity and high-energy frontiers are analysed.  
\end{abstract}
\begin{tabbing}
\> {\small Relativistic Degrees of Freedom}
\end{tabbing}

\newpage
\tableofcontents
\newpage

\section{Introduction}\label{sec:Intro}
\setcounter{equation}{0}
The existence of matter-antimatter asymmetry within the Universe, as well as the observations of the flavour mixing of neutrino species~\cite{Ahmad:2001an,Ahmad:2002jz,Fukuda:1998mi}, provide convincing evidence of physics beyond the Standard Model (SM). In recent years, the Wilkinson Microwave Anisotropy Probe~\cite{WMAP:2003ogi} and Planck observatory~\cite{Planck:2018vyg} have measured the Baryon Asymmetry of the Universe (BAU) to an unprecedented precision,
\begin{equation}
   \label{eq:etaBCMB}
    \eta_B^{\rm CMB} = (6.104 \pm 0.058)\times 10^{-10}.
\end{equation}
This value of $\eta_B$ is compatible with the one derived from Big-Bang Nucleosynthesis (BBN) bounds on the abundance of light elements~\cite{Fields:2019pfx}.

A minimal and appealing solution to the origin of the observed matter-antimatter asymmetry in the Universe offers the well-studied scenario of leptogenesis~\cite{FukYan:1986}. This is a simple framework which comfortably satisfies all three Sakharov conditions~\cite{Sakharov:1967dj}, and only requires a minimal extension of the SM to contain right-handed neutrino species, which are singlets under the $\textrm{SU}(3)_c \times \textrm{SU}(2)_L \times \textrm{U}(1)_Y$ gauge group of the SM. This minimal extension provides an additional Yukawa coupling with the left-handed neutrino species as well as a Majorana mass term, which violates the lepton number, $L$, by two units. This minimal extension not only provides a mechanism by which SM neutrinos may be massive but also ensures that this mass is small enough to fit with experimental observations through the famous {\em seesaw} mechanism~\cite{Minkowski:1977sc,Yanagida:1979as,Mohapatra:1979ia,GellMann:1980vs,Schechter:1980gr}. Moreover, the expansion of the FRW Universe leads to cooling, allowing particle species that are part of the thermal plasma to fall out of equilibrium, thus providing a mechanism for satisfying Sakharov's third condition. Finally, this framework introduces new sources of CP violation from the neutrino--Yukawa sector, as well as phenomena of heavy-neutrino mixing~\cite{Pilaftsis:1997dr,Pilaftsis:1997jf} which include coherent oscillations between neutrino flavours~\cite{Akhmedov:1998qx,Asaka:2005pn}. These effects together permit the generation of a significant lepton asymmetry, which may be rapidly converted into a baryon asymmetry through $(B+L)$-violating sphaleron transitions~\cite{KUZMIN198536}.

One issue which is commonly drawn from models of leptogenesis is that for a generic model, one would typically require singlet neutrino masses of Grand Unified Theory (GUT) scales due to the lightness of the SM neutrinos. As a result, any hopes of detection at current and future experiments are lost since the interactions between light and heavy neutrino species are suppressed by the heavy neutrino masses. However, an elegant framework by which this may be evaded is the framework of \textit{Resonant Leptogenesis} (RL)~\cite{Pilaftsis:1997jf,Pilaftsis:2003gt,Pilaftsis:2005rv}. In RL models, the CP asymmetry is enhanced through the mixing of near-degenerate heavy neutrinos, which have masses satisfying the condition
\begin{equation}
   \label{eq:mass-diff}
    \left| m_{N_\alpha} - m_{N_\beta} \right|\, \simeq\, \frac{1}{2}\Gamma_{\alpha,\beta}\;.
\end{equation}
Here, $m_{N_\alpha}$ and $\Gamma_\alpha$ are the mass and decay width of the heavy neutrino species $N_\alpha$, respectively. This framework can achieve successful leptogenesis at sub-$\TeV$ scales whilst maintaining agreement with current oscillation parameters. Consequently, RL models offer not only the possibility of generating the observed BAU but also of observing effects in near-future experiments.

In this article, we adopt a particular framework of RL with three singlet neutrino species with a mass spectrum in consecutive resonance, a structure which we refer to as {\em tri-resonant}~\cite{daSilva:2022mrx}, leading to Tri-Resonant Leptogenesis (TRL). In a TRL model, the CP asymmetry is maximised through the constructive interference of all three heavy neutrino species, allowing for greater amounts of baryon asymmetry to be generated with larger-scale Yukawa couplings. In particular, this is to be contrasted with the expectation from typical bi-resonant models where the CP asymmetry is generated through the mixing of two heavy neutrinos, with a third decoupled neutrino included to fit neutrino oscillation data. 

A notable addition to previous research on this topic is the incorporation of the non-constant nature of relativistic degrees of freedom ({\em dofs}) which we study in detail. These enter through the energy and entropy densities of the thermal plasma in the Transport Equations (TEs). It has become a common practice in the literature that at temperatures, $T$, above the electroweak (EW) scale, the {\em dofs} of the plasma are taken to be constant. As shown in a previous study~\cite{daSilva:2022mrx}, however, even small deviations which are present at this scale can have a significant effect on the generated BAU, in particular for models where the heavy neutrino mass scale is~below $100~\GeV$. In this work, we utilise a set of TEs which take into account the CP-violating effects from both mixing and heavy-neutrino flavour oscillations and show that the impact on the generated BAU from the inclusion of the EW-scale temperature variations of the {\em dofs} is the same as shown in ref.~\cite{daSilva:2022mrx}. In addition, we show that the inclusion of the {\em dofs} can induce large dependencies of BAU on the critical temperature of the sphaleron transitions, $T_{\rm sph}$, for low-scale models, in contrast with near $\TeV$ scale models where variations in $T_{\rm sph}$ have minimal effect.

Finally, we consider whether the critical phenomena discussed in~\cite{Karamitros:2022oew} may also be present in a thermal plasma. Our analysis shows that similar phenomena may be present in the approach to equilibrium. However, in a cosmological setting where the Universe cools down, these phenomena may be destroyed, with the out-of-equilibrium effects pushing the bath towards a fully mixed state. Conversely, in thermo-static scenarios (i.e.~of a thermal bath with constant temperature), it may be possible to observe oscillations in the neutrino number density as the plasma approaches its equilibrium distribution.

The layout of this article is as follows. In Section~\ref{sec:Model}, we introduce the TRL model which is based on the symmetry-motivated $\mathbb{Z}_6$ Yukawa structure. We discuss some of the CP properties which arise from this structure. In Section~\ref{sec:TEqns}, we introduce a set of coupled TEs which describe the evolution of an appreciable BAU, crucially preserving the temperature dependence of the relativistic {\em dofs} and including both the mixing of heavy neutrino species as well as coherent flavour oscillations. Section~\ref{sec:Entropy} investigates how the inclusion of the temperature dependence of the {\em dofs} impacts the generated BAU and the dependence on the critical temperature of the $(B+L)$-violating sphaleron transitions is analysed. Moreover, we examine the stability of the generated asymmetry under changes to the initial conditions. In Section~\ref{sec:Crit}, we consider the prevalence of critical scenarios within cosmological settings and discuss the TRL model as one which can produce critical phenomena. 
In this regard, we present in Appendix~\ref{App:Jordan} the mathematical structure of an $N$-level unstable quantum system of the non-diagonalizable Jordan\- form, which may also be called a critical unstable qudit. In Section~\ref{sec:Results}, we summarise our results and show numerical solutions to the TEs.\footnote{The numerical integration of the TEs and the production of the necessary interpolations is accomplished by using the modules provided by {\tt scipy}\cite{2020SciPy-NMeth}. All the figures are produced using the library {\tt matplotlib}\cite{Hunter:2007}.} The accessible parameter space generated by the TRL model is delineated, and comparisons to current and future experiments are given. Finally, in Section~\ref{sec:Concl}, we concisely present our conclusions and discuss possible extensions and future directions.

\section{The Tri-Resonant Leptogenesis Model}\label{sec:Model}
\setcounter{equation}{0}
We utilise a minimal extension of the SM through the inclusion of right-handed neutrino fields, $\nu_{R}$, which are singlets of the SM gauge group and have lepton number $L_{\nu_R} = 1$. The inclusion of these additional degrees of freedom leads to the extended Lagrangian
\begin{equation}\label{eq:Lagr}
    -\mathcal{L}^{\nu_R} = \boldsymbol{h}^{\nu}_{ij} \overline{L}_i \tilde{\Phi} \nu_{R,j} + \frac{1}{2}\overline{\nu}_{R,i}^C \lrb{\boldsymbol{m}_M}_{ij} \nu_{R,j} + {\rm H.c.} \; ,
\end{equation}
where $L_i = \lrb{\nu_{i\, L},\, e_{i\, L}}^{\sf T}$ with $i = e, \, \mu, \, \tau$ are the left-handed ${\rm SU(2)}_L$ lepton doublets, and $\tilde{\Phi} = i\sigma_2\Phi^*$ is the weak-isospin-conjugate Higgs doublet of~$\Phi$. The two flavour space matrices $\boldsymbol{h}^{\nu}$ and $\boldsymbol{m}_M$ are the neutrino Yukawa couplings and the Majorana mass matrix, respectively. We have reserved the boldface notation to indicate matrices with flavour structure. Without loss in generality, we may employ covariant flavour-space transformations
to bring the Majorana mass matrix into the diagonal form: $\boldsymbol{m}_M = {\rm diag}(m_{N_1}, \cdots, m_{N_n})$.

In the broken phase of the theory, this Lagrangian gives rise to a light neutrino mass spectrum described by the famous seesaw relation~\cite{Minkowski:1977sc,Yanagida:1979as,Mohapatra:1979ia,GellMann:1980vs,Schechter:1980gr},
\begin{equation}\label{eq:seesaw}
    \boldsymbol{m}^\nu \approx - \boldsymbol{m}_D \boldsymbol{m}_M^{-1} \boldsymbol{m}_D^{\sf T}\, ,
\end{equation}
where $\boldsymbol{m}_D = v\, \boldsymbol{h}^\nu / \sqrt{2}$ is the Dirac-neutrino mass matrix, with $v$ being the vacuum-expectation-value (VEV) of the Higgs field, and
$\boldsymbol{m}_M$ is the Majorana mass matrix describing
the spectrum of heavy neutrino species, $N_\alpha$, with $\alpha =1,2,\dots, n$. In a generic model, the seesaw relation suggests the existence of GUT-scale heavy neutrinos since the mass scale of the Majorana mass matrix is determined by
\begin{equation}
    |\!|\boldsymbol{m}_M|\!|\, \simeq\, \frac{v^2}{2|\!|\boldsymbol{m}^\nu|\!|} |\!|\boldsymbol{h}^\nu|\!|^2 \, .
\end{equation}
Here we made use of the Frobenius norm, e.g., 
\begin{equation}
    |\!| \boldsymbol{m}_M |\!|\, =\, \sqrt{\textrm{Tr}\Big(\boldsymbol{m}_M^\dagger \boldsymbol{m}_M\Big)} \, .
\end{equation}
The Lagrangian of the phenomenologically relevant $W \ell N$ charged current interaction may be written as
\begin{equation}\label{eq:ChargedCurrentInt}
    \mathcal{L}^W_{\rm int} = -\frac{g_w}{\sqrt{2}}W^-_\mu
  \overline{e}_{iL} B_{i\alpha}\gamma^\mu P_L N_\alpha + {\rm H.c.}\, ,
\end{equation}
where $P_L = \frac{1}{2}\lrb{\mathds{1} - \gamma_5}$ is the left-handed chirality projection operator. Consequently, the phenomena arising from such heavy neutrinos would be out of reach in current and foreseeable experiments, since, in the broken phase, these interactions are severely suppressed by the light-to-heavy-neutrino-mixing matrix, $B \simeq \boldsymbol{m}_D\boldsymbol{m}_M^{-1}$, where the conventions of~\cite{Pilaftsis:1991ug} were adopted. Therefore, there is great phenomenological interest in finding models where observable physics lies closer to current experimental bounds, such as at sub-$\TeV$ mass scales.

As was detailed in \cite{daSilva:2022mrx}, by selecting a symmetry-motivated flavour structure for the Yukawa, the light neutrino spectrum can be made to vanish, with the observed masses and mixing observables generated through small perturbations about the symmetric Yukawa couplings. We may parameterise the Majorana mass matrix with small mass splittings as
\begin{equation}
    \boldsymbol{m}_M\, =\, m_N \mathds{1}_N\, +\, \Delta \boldsymbol{m}_M\;. 
\end{equation}
In this expression, we use the average mass for $N$ heavy neutrino species $m_N$. At the leading order, the expression for the light neutrino mass matrix may be given by
\begin{equation}
    \boldsymbol{m}^\nu = - \frac{v^2}{2 m_N} \lrb{\boldsymbol{h}^{\nu}(\boldsymbol{h}^\nu)^{\sf T} + \mathcal{O}\lrb{\frac{\Delta \boldsymbol{m}_M}{m_N} }}\, .
\end{equation}
Hence, for singlet neutrinos with a near-degenerate mass spectrum, the light neutrino mass matrix may be made to approximately vanish by demanding that
\begin{equation}
    \boldsymbol{h}^{\nu}(\boldsymbol{h}^\nu)^{\sf T} = \boldsymbol{0}_3 \, .
\end{equation}
This motivates the symmetric Yukawa structure
\begin{equation}\label{eq:Yukawa}
    \boldsymbol{h}^\nu_0 = \begin{pmatrix}
        a & a\, \omega & a\,\omega^2\\
        b & b\,\omega & b\,\omega^2\\
        c & c\,\omega & c\,\omega^2
    \end{pmatrix}\; ,
\end{equation}
where $a,\, b, \, c \, \in \mathbb{C}$, and $\omega$ are the generators of the discrete group $\mathbb{Z}_6$. We note that due to the isomorphism $\mathbb{Z}_3 \simeq \mathbb{Z}_6 / \mathbb{Z}_2$, the generators of $\mathbb{Z}_3$ also satisfy the zero mass condition, and so they may be used to produce identical results up to a change in the sign in the CP phase. For definiteness, however, we explicitly take the generators of~$\mathbb{Z}_6$.

It is worth highlighting the following observation. Since the above symmetry-motivated  $\mathbb{Z}_6$ form appears in the flavour structure of the Dirac mass matrix, flavour covariance entails the vanishing of the tree-level mass matrix which in turn leads to the vanishing of the sum of the tree plus the one-loop contribution as well~\cite{Pilaftsis:1991ug, Pilaftsis:1998pd,
Dev:2012sg}:
\begin{align}
    \boldsymbol{h}^\nu\left(\m^{-1}_M - \frac{\alpha_w}{16\pi
  M^2_W} \, \m^{\dagger}_M \, f(\m_M\m^\dagger_M)\right)\boldsymbol{h}^{\nu\sf
  T}=\,\mathbf{0}_3\;, 
    \label{eq:one_loop_zero_mass}
\end{align}
where
\begin{align}
    f(\m_M\m^\dagger_M)=\frac{M^2_H}{\m_M\m^\dagger_M -
  M^2_H\Id_3}\ln\left(\frac{\m_M\m^\dagger_M}{M^2_H}\right) +
  \frac{3M^2_Z}{\m_M\m^\dagger_M -
  M^2_Z\Id_3}\ln\left(\frac{\m_M\m^\dagger_M}{M^2_Z}\right)\;. 
    \label{eq:loop_factor_f_def}
\end{align}
Here, $\alpha_w\equiv g_w^2/(4\pi)^2$ is the electroweak-coupling parameter, and $M_W$, $M_Z$, and $M_H$ are the  respective masses of the $W$, $Z$, and Higgs bosons. As a consequence, the symmetric structure we present gives full control over the light neutrino mass spectrum. As was described in~\cite{daSilva:2022mrx}, by judiciously rescaling the Yukawa couplings through the Majorana mass terms, it is possible to reinforce that the $\mathbb{Z}_6$ symmetry gives exactly vanishing neutrino masses to all loop orders in perturbation theory.

The $\mathbb{Z}_6$ structure not only offers naturally light neutrino masses but also has large CP-phases available for generating significant lepton asymmetry. When one calculates the CP-odd invariant quantity \cite{Pilaftsis:1997jf, Pilaftsis:2003gt, Branco:1986gr, Yu:2020gre}
\begin{equation}
    \Delta^{\rm CP} = {\rm Im} \, \textrm{Tr}\lrb{\boldsymbol{h}^{\nu\,\dagger} \boldsymbol{h}^\nu\boldsymbol{m}_M^\dagger\boldsymbol{m}_M\boldsymbol{m}_M^\dagger\boldsymbol{h}^{\nu\,\dagger} \boldsymbol{h}^{\nu\,*} \boldsymbol{m}_M} \, ,
\end{equation}
it can be shown that the resulting expression for the $\mathbb{Z}_6$ model depends only on the imaginary part of the generator, $\omega$, {\it viz.}
\begin{equation}
    \Delta^{\rm CP} = \lrb{|a|^2 + |b|^2 + |c|^2}^2 \sum_{\alpha < \beta} m_{N_\alpha} m_{N_\beta} \lrb{m_{N_\alpha}^2 - m_{N_\beta}^2} \Im{\omega^{2\lrb{\alpha-\beta}}}\, ,
\end{equation}
where $m_{N_\alpha}$ is the mass of the heavy neutrino $N_\alpha$.

In order to maximise the available CP asymmetry, we make use of this symmetric structure within the RL framework~\cite{Pilaftsis:2003gt}. This framework generates enhanced CP-violating effects through the mixing of heavy neutrino species by taking a near-degenerate heavy neutrino mass spectrum, in particular, through the contributions from the absorptive part of the wavefunction~\cite{PhysRevD.48.4609, Flanz:1994yx, Covi:1996wh, Flanz:1998kr}. Making use of the wavefunction and vertex coefficients
\begin{align}
    A_{\alpha \beta} &= \sum_{l=1}^3 \frac{\boldsymbol{h}^\nu_{l \alpha}\h_{l
                       \beta}^{\nu*}}{16\pi} =
                       \frac{(\boldsymbol{h}^{\nu\dagger}\boldsymbol{h}^\nu)^*_{\alpha\beta}}{16\pi} \,,\\ 
    V_{l \alpha} &= \sum_{k=1}^3 \sum_{\gamma \neq \alpha}
                   \frac{\boldsymbol{h}^{\nu*}_{k\alpha}\boldsymbol{h}^\nu_{k
                   \gamma}\boldsymbol{h}^\nu_{l\gamma}}{16\pi} \, f\left(
                   \frac{m^2_{N_\gamma}}{m^2_{N_\alpha}} \right) \, ,
    \label{eq:V_def}
\end{align}
a fully consistent resummation of these two wavefunction and vertex contributions is obtained through the effective Yukawa couplings~\cite{Pilaftsis:2003gt, Pilaftsis:2005rv, Deppisch:2010fr, daSilva:2022mrx}
\begin{align}
\label{eq:Eff_Yuk}
    (\bar{\boldsymbol{h}}^\nu_+)_{l\alpha} =&\; \boldsymbol{h}^\nu_{l\alpha} +
                                          iV_{l\alpha} - i
                                          \sum_{\beta,\gamma = 1}^3
                                          |\varepsilon_{\alpha\beta\gamma}|\,\boldsymbol{h}^\nu_{l\beta}\nonumber\\&\times
  \frac{m_{N_\alpha}\left(M_{\alpha\alpha\beta}+M_{\beta\beta\alpha}\right)-i
  R_{\alpha\gamma}
  \left[M_{\alpha\gamma\beta}\left(M_{\alpha\alpha\gamma}+M_{\gamma\gamma\alpha}\right)
  +
  M_{\beta\beta\gamma}\left(M_{\alpha\gamma\alpha}+M_{\gamma\alpha\gamma}\right)\right]}{m_{N_\alpha}^2-m_{N_\beta}^2
  + 2i m^2_{N_\alpha} A_{\beta\beta} + 2i\,\Im{
  R_{\alpha\gamma}}\left(m_{N_\alpha}^2 |A_{\beta\gamma}|^2 +
  m_{N_\beta} m_{N_\gamma} \Re{A_{\beta\gamma}^2} \right)}\,.
\end{align}
In the above expression, $M_{\alpha\beta\gamma} = m_{N_\alpha} A_{\beta\gamma}$, $\epsilon_{\alpha\beta\gamma}$ is the Levi-Civita anti-symmetric tensor, $f(x) = \sqrt{x}\lrsb{1-(1+x)\ln \lrb{\frac{1+x}{x}}}$ is the Fukugita-Yanagida one-loop function~\cite{FukYan:1986}, and 
\begin{equation}
    R_{\alpha\beta} \equiv \frac{m_{N_\alpha}^2}{m_{N_\alpha}^2-m_{N_\beta}^2+2i m_{N_\alpha}^2 A_{\beta\beta}}\;.
\end{equation}
The CP-conjugate effective Yukawa couplings, associated with the $N L^c \Phi^*$ interaction, and represented by $\bar{\boldsymbol{h}}^\nu_-$, may be found by using the CP-conjugate tree-level Yukawa couplings, $(\boldsymbol{h}^\nu)^*$, in~\eqref{eq:Eff_Yuk} in place of $\boldsymbol{h}^\nu$. The (rest frame) decay matrices of heavy neutrino species into $L\Phi$ and as the CP-conjugate decay to $L^c\Phi^*$ are then given by
\begin{equation}\label{eq:DecayRates}
    \Gamma_+ \equiv \Gamma(N\to L\Phi) = \frac{m_{N}}{8\pi} (\boldsymbol{h}_+^\nu)^\dagger \boldsymbol{h}^\nu_+, \quad \Gamma_- \equiv \Gamma(N\to L^c\Phi^*) = \frac{m_{N}}{8\pi} (\boldsymbol{h}_-^\nu)^\dagger \boldsymbol{h}^\nu_- \,.
\end{equation}

It has been well established in numerous works~\cite{Pilaftsis:1997jf, Pilaftsis:2003gt, Pilaftsis:2004xx} that the CP asymmetry generated through the mixing of two heavy neutrino species is maximised when the mass difference between two heavy neutrino species is of the same scale as the width
\begin{equation}
    \left|m_{N_\alpha} - m_{N_\beta} \right|\, \simeq\, \frac{1}{2}\Gamma_{\alpha, \beta} \,, 
    \tag{\ref{eq:mass-diff}}\nonumber
\end{equation}
with such mass splitting being described as {\em resonant}. It has additionally been shown in~\cite{daSilva:2022mrx} that the CP-asymmetry generated in models with three singlet neutrinos may be maximised when the heavy neutrino mass spectrum is in consecutive resonance with resonant mass splitting between not only $N_1$ and $N_2$, but also between $N_2$ and $N_3$. Consequently, we adopt this tri-resonant structure along with the $\mathbb{Z}_6$ symmetric Yukawa couplings in an effort to bound 
from above the size of the allowed light-to-heavy-neutrino mixings 
whilst maintaining agreement with the observed BAU.

\section{Transport Equations for Leptogenesis}\label{sec:TEqns}
\setcounter{equation}{0}
To determine the amount of BAU generated through TRL, we require a set of TEs to track the evolution of the number densities of the heavy neutrino species and the SM leptons.\footnote{We assume that the Higgs boson maintained thermal equilibrium with the plasma at least until the sphaleron decoupling temperature, i.e. $T_{\rm sph} > m_h$.} In this section, we lay out these TEs.

In~\cite{Sigl:1993ctk}, a general framework for the construction of quantum TEs was presented${}$, where a {\em matrix of densities} was proposed as a way to track the evolution of number densities. Such a matrix of densities was defined as
\begin{equation}
    f_{ij} = \frac{1}{V} \bvev{a^\dagger_i a_j} = \frac{1}{V}\Tr{\rho \, a^\dagger_i(\boldsymbol{p}) a_j(\boldsymbol{p})}.
\end{equation}
In the above, $V = (2\pi)^3 \delta(0)$ is the infinite $3$-volume of the coordinate space, $\rho$ is the density matrix of the thermal ensemble, and $a_i(\boldsymbol{p})$ are annihilation operators for the relevant particle species. In contrast with flavour diagonal TEs, this approach offers the ability to account for both the number densities and quantum correlations between them. From this matrix-of-densities approach, a set of coupled TEs may be derived, which capture CP-violating effects from both the oscillation between singlet neutrino flavours and mixing.

\subsection{Heavy Neutrino Transport Equations}

We follow a similar prescription to that laid out in \cite{BhupalDev:2014pfm}~\footnote{ For other out-of-equilibrium field theoretic treatments, see~\cite{Jukkala:2021sku,Kainulainen:2023ocv}.}, with the TEs for the distribution functions of the singlet neutrinos in the mass basis given by
\begin{subequations}\label{eq:TEsDistFunctions}
    \begin{equation}
        \frac{d f^N}{dt} = -i \lrsb{E^N, f^N} + \mathcal{C}^N + \lrb{\overline{\mathcal{C}}^N}^{\sf T}
    \end{equation}
    \begin{equation}
        \frac{d \overline{f}^N}{dt} = i \lrsb{E^N, \overline{f}^N} + \overline{\mathcal{C}}^N + \lrb{\mathcal{C}^N}^{\sf T}
    \end{equation}
\end{subequations}
Where $\mathcal{C}^N$ and $\overline{\mathcal{C}}^N$ are collision terms of the TEs. We consider the thermal bath in the relativistic limit, with the bath in kinetic equilibrium due to the rapid decay of heavy neutrinos $\Gamma \gg H(z)$. We also neglect quantum effects such as Bose enhancement and Fermi blocking. In addition, we do not consider the contribution of back-reaction terms to the heavy neutrino TEs. With these assumptions, the pertinent collision terms are written as
\begin{subequations}
    \begin{equation}
        \mathcal{C}^N = - \frac{1}{2} \lrcb{f^N, \Gamma_{+}(\boldsymbol{p}_N, \boldsymbol{p}_L)} + f^\phi_{\rm eq} f^L_{\rm eq} \Gamma_-(\boldsymbol{p}_N, \boldsymbol{p}_L),
    \end{equation}
    \begin{equation}
        \overline{\mathcal{C}}^N = - \frac{1}{2} \lrcb{\overline{f}^N, \Gamma_{-}(\boldsymbol{p}_N, \boldsymbol{p}_L)} + f^\phi_{\rm eq} f^L_{\rm eq} \Gamma_+(\boldsymbol{p}_N, \boldsymbol{p}_L).
    \end{equation}
\end{subequations}
With these definitions, we may derive a set of TEs for the CP-even and CP-odd linear combinations of distribution functions. However, as was extensively discussed in the literature~\cite{Kolb:1979qa, Pilaftsis:2003gt, BhupalDev:2014pfm}, we note that the subtraction of the so-called Real Intermediate States (RISs) in the $2\to 2$ scattering processes contributes a term at the same scale as the decay processes, i.e.
\begin{equation}
    \frac{d \delta\! f^N}{dt}\, \supset\, 4i f^\phi_{\rm eq} f^L_{\rm eq} \Im{\delta\Gamma(\boldsymbol{p}_N, \boldsymbol{p}_L)}.
\end{equation}
This term changes the sign of the CP-odd inverse decay terms, allowing the TEs to reach thermal equilibrium. In detail, these TEs read\footnote{ There exists a lack of consensus regarding the inclusion of the second term in equation~(\ref{eq:CPoddDistTE}). This term arises from the transport equations given in (\ref{eq:TEsDistFunctions}) with the inclusion of thermally corrected RIS subtractions in the $2\to 2$ scatterings, which encode the coherence-decoherence effects from the back-reaction of the thermal bath~\cite{BhupalDev:2014pfm}. Such terms are absent in other works, \textit{cf.}~\cite{Garbrecht:2018mrp}.}:
\begin{subequations}
    \begin{equation}
        \frac{d \underline{f}^N}{dt} = -\frac{i}{2}\lrsb{E^N, \delta\! f^N} - \frac{1}{2}\lrcb{\underline{f}^N - f^\phi_{\rm eq} f^L_{\rm eq}\mathds{1}, \Re{\Gamma_T(\boldsymbol{p}_N, \boldsymbol{p}_L)}} - \frac{i}{4}\lrcb{\delta\! f^N, \Im{\delta\Gamma(\boldsymbol{p}_N, \boldsymbol{p}_L)}},
    \end{equation}
    \begin{equation}\label{eq:CPoddDistTE}
        \frac{d \delta\! f^N}{dt} = -\frac{i}{2}\lrsb{E^N, \underline{f}^N} - i\lrcb{\underline{f}^N - f^\phi_{\rm eq} f^L_{\rm eq}\mathds{1}, \Im{\delta\Gamma(\boldsymbol{p}_N, \boldsymbol{p}_L)}} - \frac{1}{2}\lrcb{\delta\! f^N, \Re{\Gamma_T (\boldsymbol{p}_N, \boldsymbol{p}_L)}}.
    \end{equation}
\end{subequations}
In the above expressions we have defined the CP-even and CP-odd components of the distribution functions
\begin{equation}
    \underline{f}^N = \frac{1}{2}\lrb{f^N + \overline{f}^N} \, , \qquad \delta f^N = f^N - \overline{f}^N \, , 
\end{equation}
as well as the CP-even and CP-odd decay terms:
\begin{subequations}
    \begin{equation}
        \Gamma_T(\boldsymbol{p}_N, \boldsymbol{p}_L) = \Gamma_+(\boldsymbol{p}_N, \boldsymbol{p}_L) + \Gamma_-(\boldsymbol{p}_N, \boldsymbol{p}_L) \, ,
    \end{equation}
    \begin{equation}
        \delta\Gamma(\boldsymbol{p}_N, \boldsymbol{p}_L) = \Gamma_+(\boldsymbol{p}_N, \boldsymbol{p}_L) - \Gamma_-(\boldsymbol{p}_N, \boldsymbol{p}_L) \, .
    \end{equation}
\end{subequations}
We identify the equilibrium distribution function as the distribution which makes the RHS of both of these TEs vanish, and we then see that the equilibrium distribution of the heavy neutrino species is given by
\begin{equation}\label{eq:EqDist}
    \underline{f}^N_{\rm eq} = f^\phi_{\rm eq} f^L_{\rm eq}\mathds{1} = \exp\lrsb{-\frac{E^\phi + E^L}{T}} \mathds{1},\quad \delta\! f^N_{\rm eq} = 0.
\end{equation}
To integrate these TEs, we may utilise the kinetic equilibrium assumption,
\begin{equation}
    \underline{f}^N \approx \frac{\underline{n}^N}{n^N_{\rm eq}}f^N_{\rm eq} \, \quad \delta\!f^N \approx \frac{\delta n^N}{n^N_{\rm eq}}f^N_{\rm eq},
\end{equation}
which has been discussed in the literature~\cite{Kolb:1979qa, Basboll:2006yx, Garayoa:2009my, Hahn-Woernle:2009jyb}, where it was shown that only small deviations are to be expected. Making use of the equilibrium distribution given in equation~(\ref{eq:EqDist}), we find the corresponding equations for the number density of the neutrino species,
\begin{subequations}\label{eq:TEsTime}
    \begin{equation}
        \frac{d \underline{n}^N}{dt} = -\frac{i}{2}\lrsb{\mathcal{E}^N, \delta n^N} - \frac{1}{2}\lrcb{\frac{\underline{n}^N}{n^N_{\rm eq}} - \mathds{1}, \Re{\gamma_{L\Phi}^N}} - \frac{i}{4n^N_{\rm eq}}\lrcb{\delta n^N, \Im{\delta\gamma}},
    \end{equation}
    \begin{equation}
        \frac{d \delta n^N}{dt} = -2i\lrsb{\mathcal{E}^N, \underline{n}^N} - i\lrcb{\frac{\underline{n}^N}{n^N_{\rm eq}} - \mathds{1}, \Im{\delta\gamma}} - \frac{1}{2 n^N_{\rm eq}}\lrcb{\delta n^N, \Re{\gamma_{L\Phi}^N }}.
    \end{equation}
\end{subequations}

In these TEs, we have defined the following thermally averaged quantities:
\begin{align}
    \mathcal{E}^N &= \frac{1}{n^N}\int_{\mathbf{p}} \, E^N(\mathbf{p}) f^N_{\rm eq}(\mathbf{p}) = \frac{K_1(z)}{K_2(z)} \boldsymbol{m}_M \, ,\\
    \gamma &= \Gamma \int_{NL\Phi} \, \frac{16\pi (p_N \cdot p_L)}{m_N} f^\phi_{\rm eq} f^L_{\rm eq} = \frac{m_N^3}{2\pi^2}\frac{K_1(z)}{z} \Gamma\, ,
\end{align}
where the integral labels indicate over which Lorentz-invariant-phase-space the integral takes place, the rest-frame decay matrices, $\Gamma_{\pm}$, are calculated in equation~(\ref{eq:DecayRates}), and $K_n(z)$ are modified Bessel functions of the second kind.

The terms which enter the TEs~\eqs{eq:TEsTime} are then written as
\begin{align}
    \gamma^N_{L\Phi} = \gamma(N\to L\Phi) + \gamma(N \to L^c \Phi^\dagger) = \frac{m_N^3}{2\pi^2}\frac{K_1(z)}{z} \lrb{\Gamma_+ + \Gamma_-},
\end{align}
\begin{align}
    \delta\gamma = \gamma(N\to L\Phi) - \gamma(N \to L^c \Phi^\dagger) = \frac{m_N^3}{2\pi^2}\frac{K_1(z)}{z} \lrb{\Gamma_+ - \Gamma_-}\, .
\end{align}

We would like to write the TEs given in equation~(\ref{eq:TEsTime}) in a way which explicitly accounts for the expansion of the Universe and consequential phenomena. Following the conventions given in~\cite{Pilaftsis:2005rv}, we introduce the dimensionless run parameter $z=m_N/T$ and normalise the particle number density to the photon number density
\begin{equation}
    \eta^i (z) = \frac{n^i(z)}{n^\gamma(z)}
\end{equation}
\begin{equation}
    n^\gamma(z) = \frac{2\zeta(3)}{\pi^2}T^3 = \frac{2\zeta(3)}{\pi^2} \lrb{\frac{m_N}{z}}^3,
\end{equation}
where $\zeta(3)=1.202$ is Ap\'{e}ry's constant.

From the TEs given in equation~(\ref{eq:TEsTime}), it can be seen that it may be convenient to express the TEs in terms of CP-even and CP-odd departure-from-equilibrium matrices, which we define as
\begin{equation}\label{eq:DFEMats}
    \Delta(z) = \frac{\eta^N(z)}{\eta^N_{\rm eq}(z)} - \mathds{1}, \quad \delta(z) = \frac{\delta\eta^N(z)}{\eta^N_{\rm eq}(z)},
\end{equation}
respectively. Here, we use the approximate expression for the equilibrium value
\begin{equation}
    \eta_{\rm eq}^N(z) = \frac{z^2}{2\zeta(3)} K_2(z) \, .
\end{equation}

With all these definitions in place, we may find a set of TEs for the departure-from-equilibrium matrices, which crucially preserve the effects from changing {\em dofs}, which appear through the energy and entropy density of the thermal plasma
\begin{equation}
    \rho(T) = \frac{\pi^2}{30} g_{\rm eff}(T) T^4, \quad s(T) = \frac{2\pi^2}{45} h_{\rm eff}(T) T^3,
\end{equation}
In the TEs, the variations in the {\em dofs} are accounted for through the quantity
\begin{equation}\label{eq:deltah}
    \delta_h = 1 - \frac{1}{3}\frac{d \ln h_{\rm eff} (z)}{d \ln z}.
\end{equation}
The TEs are then written as
\begin{subequations}\label{eq:TEslnz}
    \begin{align}
        \frac{d \Delta}{d \ln z} &= \frac{\delta_h}{H(z)} \lrb{-\frac{i}{2} \lrsb{\mathcal{E}^N, \delta} - \frac{1}{2n^\gamma \, \eta^N_{\rm eq}} \lrcb{\Delta, \Re{\gamma^N_{L\Phi}}} - \frac{i}{4n^\gamma \, \eta^N_{\rm eq}}\lrcb{\delta , \Im{\delta\gamma}}}\nonumber\\
        &+\lrb{3(1 - \delta_h) - \frac{d \ln \eta^N_{\rm eq}(z)}{d \ln z}} \lrb{\Delta + \mathds{1}} \, ,
    \end{align}
    \begin{align}
        \frac{d \delta}{d \ln z} &= \frac{\delta_h}{H(z)} \lrb{-2i \lrsb{\mathcal{E}^N, \Delta} - \frac{i}{n^\gamma \, \eta^N_{\rm eq}} \lrcb{\Delta, \Im{\delta\gamma}} - \frac{1}{2n^\gamma \, \eta^N_{\rm eq}}\lrcb{\delta , \Re{\gamma^N_{L\Phi}}}}\nonumber\\
        &+\lrb{3(1 - \delta_h) - \frac{d \ln \eta^N_{\rm eq}(z)}{d \ln z}} \delta \, .
    \end{align}
\end{subequations}
In contrast with the majority of the TEs that appeared in the literature, the respective equations displayed here explicitly show the distinct phenomena that may impact the evolution of the heavy neutrino number densities, particularly the relativistic {\em dofs} and the changes in the equilibrium, $\eta_{\rm eq}^B$. Finally, in our evaluations of the two differential equations in~\eqref{eq:TEslnz}, we make use of the well-known expression for the Hubble parameter
\begin{equation}
    H(z) = \sqrt{\frac{4\pi^3 g_{\rm eff}(z)}{45}} \frac{m_N^2}{M_{\rm Pl}} \frac{1}{z^2}\, ,
\end{equation}
where $M_{\rm Pl} \simeq 1.221\times 10^{19} \: \GeV$ is the Planck mass.

\subsection{Lepton Asymmetry Transport Equation}

As in the case of the heavy neutrino TEs, we identify the {\em flavour-covariant} TE for the lepton asymmetry in the manner described in~\cite{BhupalDev:2014pfm}. However, we additionally include the extra contributions appearing from the variations in the {\em dofs}. The TEs we use contain not only decay and inverse decay terms but also relevant $2 \to 2$ scatterings. Moreover, in order to avoid double counting of the decay processes, a proper subtraction of the RIS contributions is performed. The TE for the lepton asymmetry has been written in a manner such that collision decay terms do not appear, as outlined in~\cite{Deppisch:2010fr, BhupalDev:2014pfm}. Consequently, the TE does not contain any negative contributions to the collision terms arising from the RIS subtraction of the $\Delta L =2$ interactions~\cite{Ala-Mattinen:2023rbm}.

Since we are primarily interested in the evolution at temperatures higher than that of the critical temperature of the EW phase transition, the SM leptons are massless apart from thermally generated masses, and as a result, we neglect the oscillation of SM lepton species. We assume that the sum of lepton number density is close to thermal equilibrium, and so $\eta^L + \overline{\eta}^L \simeq 2\eta_{\rm eq}^L \, \mathds{1}$. The flavour-covariant TE for the lepton asymmetry is then written as
\begin{align}
   \label{eq:TEdetaL}
    \frac{d \lrsb{\delta\eta^L}_{lm}}{d \ln z} &= \frac{\delta_h}{H(z) n^\gamma} \left(\lrsb{\delta\gamma}_{lm\alpha\beta}\Delta_{\beta\alpha} +\frac{1}{2}\lrsb{\gamma^N_{L\Phi}}_{lm\alpha\beta} \delta_{\beta\alpha} - \frac{1}{3}\lrcb{\delta\eta^L, \gamma_{L\Phi}^{L\Phi} + \gamma^{L\Phi}_{L^c\Phi^\dagger}}_{lm} \right.\nonumber\\
    &\; \left. -\frac{2}{3}\lrsb{\delta\eta^L}_{kn}\lrb{\lrsb{\gamma^{L\Phi}_{L^c\Phi^\dagger}}_{nklm}- \lrsb{\gamma_{L\Phi}^{L\Phi}}_{nklm}} - \frac{2}{3}\lrcb{\delta\eta^L, \gamma_{\rm dec}}_{lm} + \lrsb{\delta\gamma_{\rm dec}^{\rm back}}_{lm} \right) \nonumber\\ &\; + 3(1 - \delta_h) \, \delta\eta^L\, .
\end{align}
On the RHS of~\eqref{eq:TEdetaL}, the first two terms contain the decay contribution to the TE, with the rank-4 decay terms defined as
\begin{subequations}
    \begin{align}
    \lrsb{\gamma_{L\Phi}^N}_{lm\alpha\beta} &= \frac{m_N^3}{2\pi^2}\frac{K_1(z)}{z} \frac{m_N}{8\pi} \lrb{\lrb{\boldsymbol{h}^\nu_+}_{\alpha l} (\boldsymbol{h}^\nu_+)^\dagger_{\beta m} + \lrb{\boldsymbol{h}^\nu_-}_{\alpha l} (\boldsymbol{h}^\nu_-)^\dagger_{\beta m}} \, ,\\
    \lrsb{\delta\gamma}_{lm\alpha\beta} &= \frac{m_N^3}{2\pi^2}\frac{K_1(z)}{z} \frac{m_N}{8\pi} \lrb{\lrb{\boldsymbol{h}^\nu_+}_{\alpha l} (\boldsymbol{h}^\nu_+)^\dagger_{\beta m} - \lrb{\boldsymbol{h}^\nu_-}_{\alpha l} (\boldsymbol{h}^\nu_-)^\dagger_{\beta m}} \, .
\end{align}
\end{subequations}
The remaining terms contribute to the $2\to 2$ scattering. The rank-4 scattering terms corresponding to $L\Phi \to L\Phi$ and $L\Phi \to L^c\Phi^\dagger$ are defined as
\begin{multline}
   \lrsb{\gamma_{L\Phi}^{L\Phi}}_{nklm} = \\
   \sum_{\alpha \beta} \,
  \frac{2\Big[(\gamma^N_{L \Phi})_{\alpha\alpha} \,
  +\, (\gamma^N_{L \Phi})_{\beta\beta}\Big]}
  {\left(1 \,-\, 2 i \, \frac{m_{N_\alpha} - m_{N_\beta}}
    {\Gamma_{\alpha} + \Gamma_{\beta}} \right)}
  \frac{ \lrb{\boldsymbol{h}^{\nu}_-}_{l\alpha}^* \lrb{\boldsymbol{h}^\nu_+}_{k\alpha}\lrb{\boldsymbol{h}^\nu_-}_{m\beta} \lrb{\boldsymbol{h}^\nu_+}_{n\beta}^* +  \lrb{\boldsymbol{h}^{\nu}_+}_{l\beta}^* \lrb{\boldsymbol{h}^\nu_-}_{k\beta}\lrb{\boldsymbol{h}^\nu_+}_{m\alpha} \lrb{\boldsymbol{h}^\nu_-}_{n\alpha}^*}
  {\Big[(\boldsymbol{h}^{\nu,\dagger}_+ \, 
    \boldsymbol{h}_+^\nu)_{\alpha \alpha}  \,
    + \, (\boldsymbol{h}^{\nu,\dagger}_- \,
    \boldsymbol{h}_-^\nu)_{\alpha \alpha} \,
    + \, (\boldsymbol{h}^{\nu,\dagger}_+ \, 
    \boldsymbol{h}_+^\nu)_{\beta \beta}  \, 
    + \, (\boldsymbol{h}^{\nu,\dagger}_- \,
    \boldsymbol{h}_-^\nu)_{\beta \beta}\Big]^2} \, ,
\end{multline}
\begin{multline}
   \lrsb{\gamma_{L^c\Phi^\dagger}^{L\Phi}}_{nklm} = \\
   \sum_{\alpha \beta} \,
  \frac{2\Big[(\gamma^N_{L \Phi})_{\alpha\alpha} \,
  +\, (\gamma^N_{L \Phi})_{\beta\beta}\Big]}
  {\left(1 \,-\, 2 i \, \frac{m_{N_\alpha} - m_{N_\beta}}
    {\Gamma_{\alpha} + \Gamma_{\beta}} \right)}
  \frac{ \lrb{\boldsymbol{h}^{\nu}_+}_{l\beta}^* \lrb{\boldsymbol{h}^\nu_+}_{k\beta}^*\lrb{\boldsymbol{h}^\nu_+}_{m\alpha} \lrb{\boldsymbol{h}^\nu_+}_{n\alpha} + \lrb{\boldsymbol{h}^{\nu}_-}_{l\alpha}^* \lrb{\boldsymbol{h}^\nu_-}_{k\alpha}^*\lrb{\boldsymbol{h}^\nu_-}_{m\beta} \lrb{\boldsymbol{h}^\nu_-}_{n\beta}}
  {\Big[(\boldsymbol{h}^{\nu,\dagger}_+ \, 
    \boldsymbol{h}_+^\nu)_{\alpha \alpha}  \,
    + \, (\boldsymbol{h}^{\nu,\dagger}_- \,
    \boldsymbol{h}_-^\nu)_{\alpha \alpha} \,
    + \, (\boldsymbol{h}^{\nu,\dagger}_+ \, 
    \boldsymbol{h}_+^\nu)_{\beta \beta}  \, 
    + \, (\boldsymbol{h}^{\nu,\dagger}_- \,
    \boldsymbol{h}_-^\nu)_{\beta \beta}\Big]^2} \, ,
\end{multline}
with the scattering matrices defined through a contraction over the final two indices
\begin{equation}
    \lrsb{\gamma_{L\Phi}^{L\Phi}}_{lm} = \sum_k \lrsb{\gamma_{L\Phi}^{L\Phi}}_{lmkk}, \quad \lrsb{\gamma_{L^c\Phi^\dagger}^{L\Phi}}_{lm} =  \sum_k \lrsb{\gamma_{L^c\Phi^\dagger}^{L\Phi}}_{lmkk} \, .
\end{equation}
Finally, the interactions with SM particles are accounted for in the decoherence matrices $\gamma_{\rm dec}$ and $\delta\gamma_{\rm dec}^{\rm back}$. The former of these is a CP-even contribution, with $\gamma_{\rm dec} = \Gamma^L n_{\rm eq}^L$, and $\Gamma^L = {\rm diag}(\Gamma_\ell^L)$, where
\begin{equation}
    \Gamma^L_\ell(T) \simeq \lrb{3.8 \times 10^{-3}} \lrsb{\boldsymbol{h}^L_\ell}^2 \lrsb{(-1.1 + 3.0\, x) + 1.0 + \lrsb{\boldsymbol{h}^Q_t}^2(0.6-0.1)x} T \, ,
\end{equation}
as calculated in~\cite{Cline:1993bd}. In this expression, $\boldsymbol{h}^L_{\ell}$ are the Yukawa couplings for the SM leptons, $\boldsymbol{h}^Q_{t}$ is the Yukawa coupling for the top quark, $x = z\, M_\Phi(z)/ m_N$, and $M_\Phi(z)$ is the thermal Higgs mass. The CP-odd contribution may be calculated from the CP-even term as
\begin{equation}
    \lrsb{\gamma_{\rm dec}^{\rm back}}_{ll} = \lrsb{\gamma_{\rm dec}}_{ll} \frac{\delta \eta^L_{ll}}{\eta^L_{\rm eq}}\, ,
\end{equation}
where the repeated index $l$ is not summed over.

The total lepton asymmetry is easily extracted from $\delta\eta^L$ through its trace. Part of this lepton asymmetry is then re-processed into a baryon asymmetry through $(B+L)$-violating sphaleron transitions, which become exponentially suppressed at the sphaleron critical temperature $T_{\rm sph} \simeq 132\:\GeV$~\cite{DOnofrio:2014rug}. In~\cite{PhysRevD.42.3344}, the conversion factor for leptons-to-baryons is calculated to be
\begin{equation}
    \eta_B \simeq - \frac{28}{51} \: {\rm Tr} \: \delta\eta^L \, .
\end{equation}
Finally, in order to compare the generated BAU with the observed value, we should take into account the dilution of $\eta_B$ due to the expansion of the Universe. At temperatures below $T_{\rm sph}$, we assume that the total number of baryons is fixed, but the number density may be washed out due to the expansion of the Universe. Assuming that there are no considerable entropy-releasing processes between the sphaleron temperature and the recombination temperature, $T_{\rm rec}$, we may use entropy conservation to estimate
\begin{equation}\label{eq:dilution}
    \eta_B^{\rm CMB} = \frac{h_{\rm eff}(T_{\rm rec})}{h_{\rm eff}(T_{\rm sph})} \, \eta_B(T_{\rm sph}),
\end{equation}
with the overall conversion factor found to be around $1/27$~\cite{Pilaftsis:2003gt, Buchmuller:2004nz}. Therefore, the BAU, as measured at the recombination epoch, may be extracted from the TEs by
\begin{equation}
    \eta_B^{\rm CMB} = -\frac{1}{27}\frac{28}{51} \, {\rm Tr} \,\delta \eta^L \, .
\end{equation}

\section{Impact of the Relativistic Degrees of Freedom}\label{sec:Entropy}
\setcounter{equation}{0}
In order to estimate the reliability of our results in different regions, in this section we investigate how the results for the generated BAU change once the temperature dependence of the relativistic {\em dofs} is included. As  shown in the literature \cite{Hindmarsh:2005ix, Gondolo:1990dk}, the contributions to the effective {\em dofs} are calculated through the energy and entropy densities of the Universe
\begin{equation}
    g_{\rm eff}(T)= \frac{30}{\pi^2} \frac{\rho(T)}{T^4}\;, \qquad  h_{\rm eff}(T)= \frac{45}{2\pi^2} \frac{s(T)}{T^3}\; . 
\end{equation}
The inclusion of lattice~\cite{Karsch:2000ps} and perturbative~\cite{Kajantie:2002wa} QCD effects gives additional contributions to the equation of state for the plasma. This results in an equation of state that deviates from the ideal gas assumption in a relevant manner~\cite{Hindmarsh:2005ix} and, consequently, small variations in the {\em dofs} are present at temperatures above $100\:\GeV$.

In the TEs, the variations in the {\em dofs} are accounted for in three places: (i)~the Hubble rate,~$H$, (ii)~as a global factor of $\delta_h$ on the commutator and collision terms, and (iii)~as an additional term proportional to $(\delta_h - 1)$. It is worth noting that these three factors do not affect our TEs in the same way. When the variations to the {\em dofs} enter into the Hubble rate, the~$T^2$ scaling of $H$ dominates over the small variations in~$g_{\rm eff}(T)$. Similarly, the global factor~$\delta_h$ is overshadowed by the temperature dependence of the thermally averaged energy and decay matrices. However, the final term proportional to~$(\delta_h - 1)$ modifies the TEs in a significant way by introducing additional dynamics. In particular, this term provides an extra way in which the number density may be drawn away from equilibrium. With this in mind, we study the dynamics of the system in two initial-condition scenarios: (i)~the case where the bath of heavy neutrinos begins in equilibrium, (ii)~the case with a vanishing number density of heavy-neutrinos as the initial condition of the bath.

\subsection{Equilibrium Initial Conditions}\label{sec:DofsEqu}

In the early evolution, we assume that thermal out-of-equilibrium effects generate the initial heavy neutrino asymmetry. Therefore, we take the initial condition $\Delta(z_0) = \delta(z_0) = 0$ when~$z_0 \ll 1$. Neglecting neutrino flavour oscillations and the collision terms, we may study the out-of-equilibrium dynamics alone. We consider the equation
\begin{equation}\label{eq:dDeltadlnz_EquilibriumInitialConditions_Dd0}
    \frac{d \Delta(z) }{d \ln z} = -\frac{d \ln \eta^N_{\rm eq}(z)}{d \ln z} \lrb{\Delta + \mathds{1}} + 3(1 - \delta_h) \lrb{\Delta + \mathds{1}} \, .
\end{equation}
The first term in~\eqs{eq:dDeltadlnz_EquilibriumInitialConditions_Dd0} gives the dynamics from the change in the equilibrium number density as the Universe is cooling down, while the second term describes the inclusion of changes in the entropy~{\em dofs}. Taking the equilibrium initial condition, it may be shown that this equation has the solution
\begin{equation}\label{eq:OOE}
    \Delta(z)\, =\, \Bigg(\frac{h_{\rm eff}(z)}{h_{\rm eff}(z_0)}\frac{\eta^N_{\rm eq}(z_0)}{\eta^N_{\rm eq}(z)} - 1 \Bigg)\,\mathds{1}\, =\, \frac{\eta^N_{\rm eq}(z_0)}{\eta^N_{\rm eq}(z)}\, \mathcal{R}\, \mathds{1} \, .
\end{equation}
In the above, we have defined the dimensionless parameter
\begin{equation}
    \mathcal{R}\: =\: \frac{h_{\rm eff}(z)}{h_{\rm eff}(z_0)}\, -\, \frac{\eta^{N}_{\rm eq}(z)}{\eta^{N}_{\rm eq}(z_0)} \, ,
\end{equation}
which will allow us to showcase better the relevant information as $z>1$. In fact, the value of~$\mathcal{R}$ provides important information on the dominant source of out-of-equilibrium dynamics. From~(\ref{eq:OOE}), it can be seen that the sign of $\Delta$ is equal to that of $\mathcal{R}$. Moreover, from the definition of $\mathcal{R}$, we can determine that when $\mathcal{R} < 0$, the change in the entropy {\em dofs} becomes more significant than the reduction in the equilibrium number density, whereas the converse is true for $\mathcal{R} > 0$. Finally, since both $h_{\rm eff}$ and $\eta^N_{\rm eq}$ decrease monotonically with $z$, the value of $\mathcal{R}$ must lie in the range $\lrsb{-1,1}$. The extreme values represent scenarios with either constant {\em dofs} and a significant reduction in the equilibrium distribution $\lrb{\mathcal{R} \to 1}$, or a scenario with minimal losses in the equilibrium number density but with a large reduction in the relativistic {\em dofs}~$\lrb{\mathcal{R} \to -1}$. In order to acquire a better understanding of the early behaviour of the number density, thanks to the linearity of the TEs, we may utilise $\mathcal{R}$ as a measure of the interplay between the two sources of out-of-equilibrium dynamics.

\begin{figure}[t]
    \centering
    \begin{subfigure}{0.49\linewidth}
        \centering
        \includegraphics[width=\linewidth]{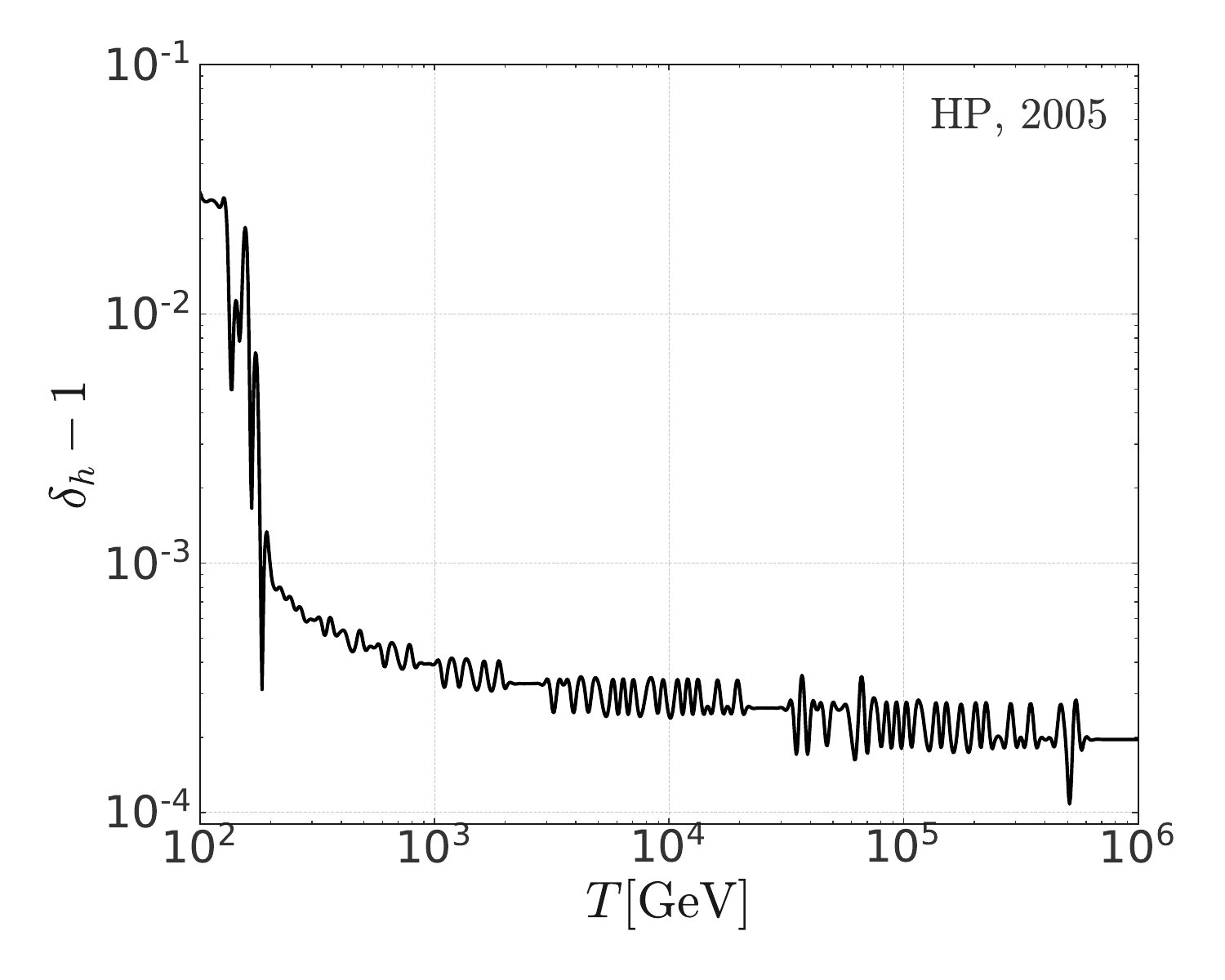}
        \caption{\empty}
        \label{fig:dH}
    \end{subfigure}
    \begin{subfigure}{0.49\linewidth}
        \centering
        \includegraphics[width=\linewidth]{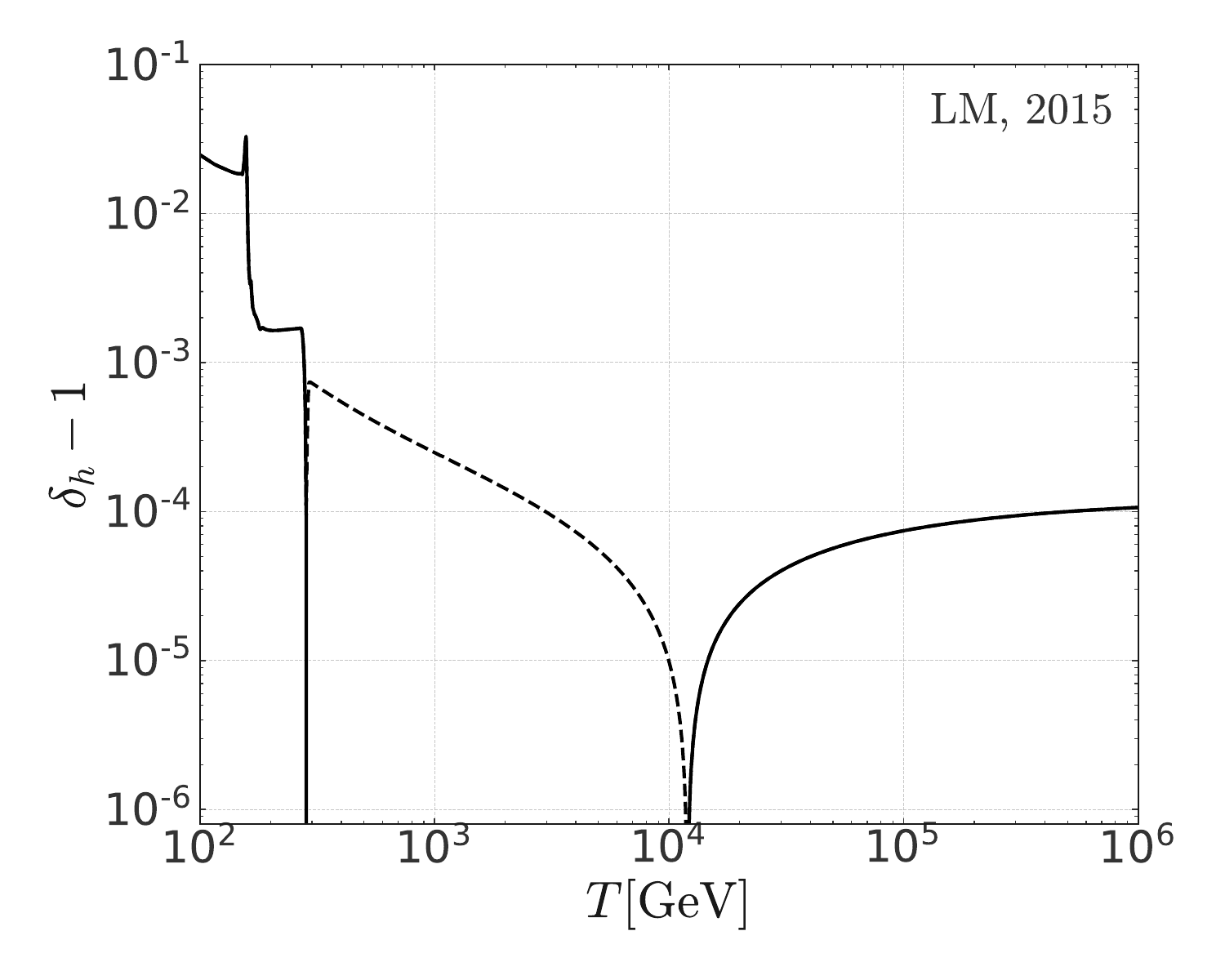}
        \caption{\empty}
        \label{fig:dHLaine}
    \end{subfigure}
    \caption{The deviation away from unity of the $\delta_h$ parameter. Observe that at temperatures well above the electroweak mass scale, the relativistic {\em dofs} are almost constant, $\delta_h \to 1$. As the temperature approaches this scale, SM particles (e.g. the top quark) start to decouple, and $\delta_h$ decreases. In this figure, we consider two datasets. The left panel uses data extracted from EOS C by Hindmarsh and Philipsen~\cite{Hindmarsh:2005ix} (HP, 2005), and the right panel shows the latest data from Laine and Meyer~\cite{Laine:2015kra} (LM, 2015). Solid lines express positive values  of $\delta_h-1$, whereas dashed lines indicate negative values.}
    \label{fig:DegreesAna}
\end{figure}

In Figure~\ref{fig:DegreesAna}, we display the effect of the {\em dofs} on the TEs. Specifically, this figure shows the value of $\delta_h - 1$ for temperatures $T>100~\GeV$. If the {\em dofs} were constant, we would expect this term to vanish. As can be seen in these figures, the value of $\delta_h - 1$, albeit small, is non-vanishing, even as the temperature increases by multiple orders of magnitude. We consider two independent parameterisations of the {\em dofs} using data from Hindmarsh and Philipsen~\cite{Hindmarsh:2005ix} (${\rm HP}, \, 2005$) and from Laine and Meyer~\cite{Laine:2015kra} (${\rm LM}, \, 2015$). Both of these data sets are generated using lattice techniques, with the former data matched to two-loop perturbative computations and the latter using a dimensionally reduced effective field theory approach to compute corrections to thermodynamic potentials. In this context, one may notice that Figures~\ref{fig:dH} and \ref{fig:dHLaine} show drastically different predictions for $\delta_h -1$ between the two parameterisations when~${10^2 \: {\rm GeV}\lesssim T \lesssim 10^4 \: {\rm GeV}}$. To explore this difference, however, a dedicated study, which goes beyond the scope of this work, may be necessary. In Figure~\ref{fig:dH}, the temperature dependence of the {\em dofs} is extracted from a discrete set (EOS C from \cite{Hindmarsh:2005ix}). Consequently, the value of $\delta_h$ at each point was necessarily numerically estimated. The functional form is then generated from a cubic spline of these points, and we employ the derivative of this spline to find $\delta_h$. As such, some accuracy may be lost, particularly in regions where the data changes rapidly. In contrast, figure~\ref{fig:dHLaine} is extracted from the data points provided in~\cite{Laine:2015kra} using the analytic relationship
\begin{equation}
    \delta_h - 1 \: = \: \frac{c}{3s}\, ,
\end{equation}
where $c$ is the heat capacity of the thermal bath, and $s$ is the entropy density. A cubic spline is then taken of these data points to produce the continuous behaviour shown in Figure~\ref{fig:dHLaine}.

\begin{figure}
    \centering
    \begin{subfigure}{0.49\linewidth}
        \centering
        \includegraphics[width=\linewidth]{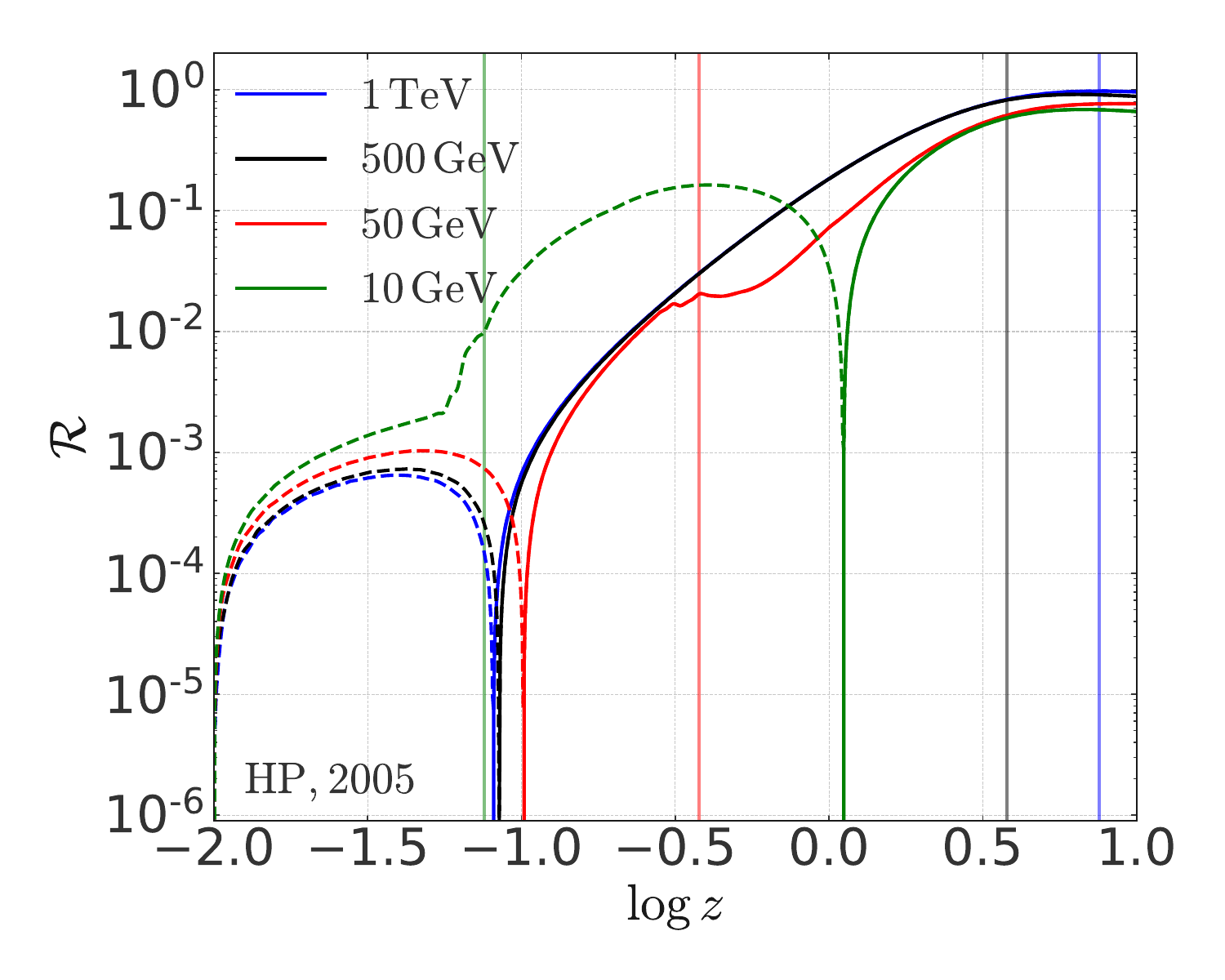}
        \caption{\empty}
        \label{fig:EntropyEffs}
    \end{subfigure}
    \begin{subfigure}{0.49\linewidth}
        \centering
        \includegraphics[width=\linewidth]{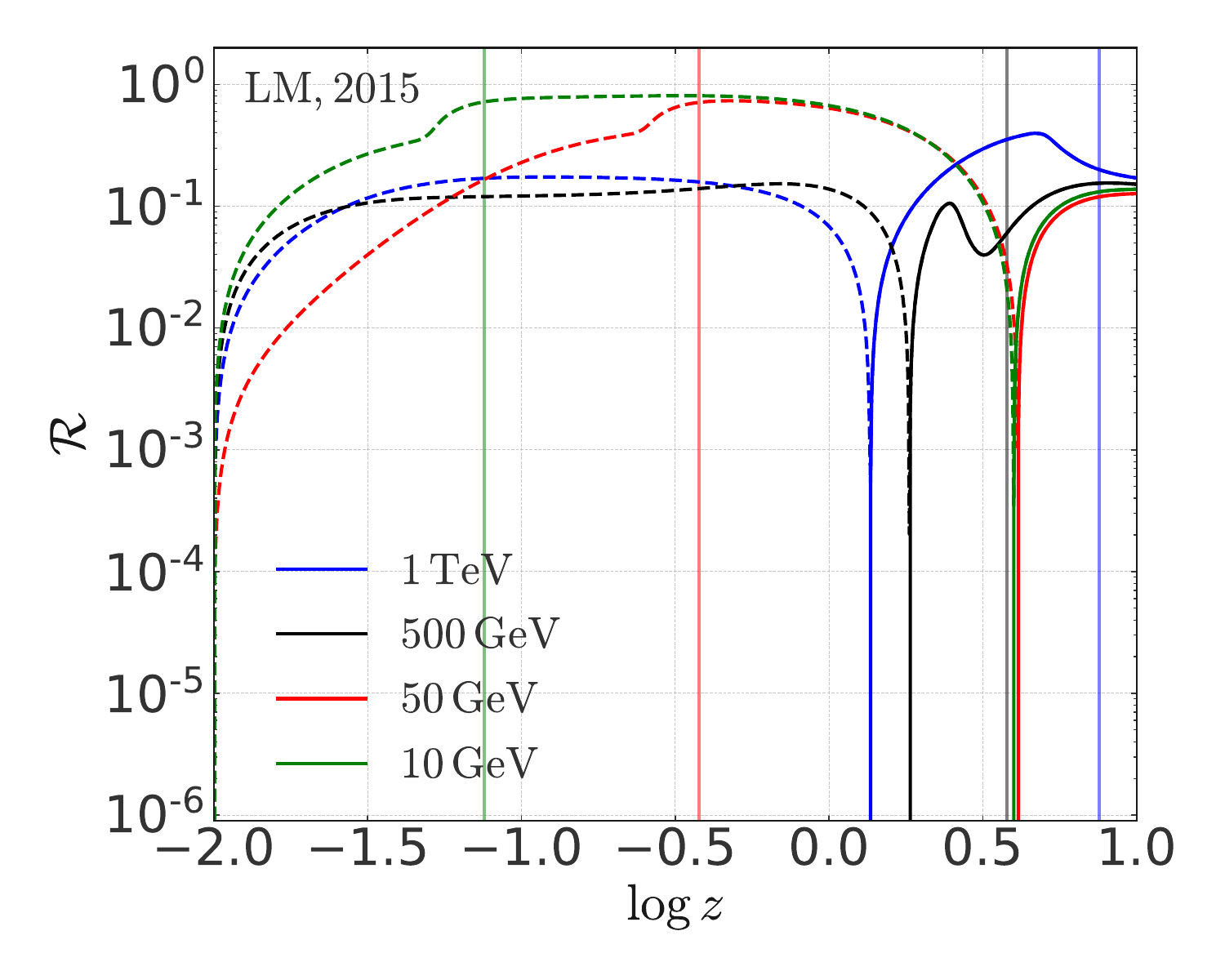}
        \caption{\empty}
        \label{fig:EntropyEffsLaine}
    \end{subfigure}
    \caption{This figure indicates the dominant out-of-equilibrium effect in the TEs. Dashed lines show where the washout of the number density due to changes in the relativistic {\em dofs} is the dominant contribution $(\mathcal{R} < 0)$. Solid lines indicate where the reduction in the equilibrium number density is the dominant effect $(\mathcal{R} > 0)$. Vertical lines show the critical temperature of the sphaleron transitions. The left panel is produced using data from Hindmarsh and Philipsen~\cite{Hindmarsh:2005ix} (HP, 2005), and the right panel utilises the data given by Laine and Meyer~\cite{Laine:2015kra} (LM, 2015).}
    \label{fig:EntropyEffects}
\end{figure}

In Figure~\ref{fig:EntropyEffects}, we show the evaluation of $\mathcal{R}$ over the range of $z$ pertinent to leptogenesis. We show four Majorana mass scales to represent scenarios at a very low scale and others above the electroweak scale. In the same figure, solid lines represent positive values of~$\mathcal{R}$, and dashed lines represent negative values. Also included are vertical lines in representative colours to show when the sphalerons become suppressed. As can be seen in Figure~\ref{fig:EntropyEffects}, for scenarios with Majorana masses well above the electroweak scale, we expect the variation in the {\em dofs} to have minimal influence on the dynamics. The simple reason is that at the sphaleron critical temperature, the value of $\mathcal{R}$ lies close to positive unity, i.e.~$\mathcal{R} \simeq +1$. This should be contrasted with low-scale scenarios, where $\mathcal{R}$ takes values closer to zero and may also be negative, i.e.~$\mathcal{R} \lesssim 0$. The latter implies a greater interplay between the two sources of out-of-equilibrium dynamics. For $10\, \GeV$ Majorana neutrinos, we see that the freeze-out of entropy {\em dofs} is the primary out-of-equilibrium effect, which is expected to be reflected in the evolution of the number densities. If the Majorana mass scale is around $50\:\GeV$, the variations in the entropy {\em dofs} have become the subdominant phenomena, but they still remain non-negligible. Therefore, we expect that the evolution of the neutrino number density has complex dynamics at this scale. Importantly, the $\mathcal{R}$ parameter we consider provides predictions for regions where the inclusion of {\em dofs} will be significant, in a way which does not rely upon computationally evaluated values of $\delta_h$.

The importance of variations within the relativistic degrees of freedom is not entirely unexpected. In the absence of interactions, the number density dilutes inversely proportional to the co-moving volume, i.e. $n^N \propto a^{-3}$. However, it is also known that the temperature dependence of the number density scales as $n^N \propto T^{3}$. In the absence of entropy-releasing transitions, conservation of the total entropy may be used to find the condition
\begin{equation}
    h_{\rm eff}(T) \, a^3 \, T^3 = {\rm const.}
\end{equation}
In the standard paradigm, with constant degrees of freedom, this condition implies an inverse proportionality between the scale factor $a$ and the temperature of the bath, $T$, reconciling the two scenarios outlined above. However, when variations in the relativistic degrees of freedom are included, the scale factor is related to the temperature through
\begin{equation}
    a \: \propto \:  \frac{1}{T \, h_{\rm eff}^{1/3}(T)} \; .
\end{equation}
As a consequence, extra contributions are present, which provide another mechanism by which the number density may be moved out of chemical equilibrium.

\subsection{Vanishing Initial Heavy-Neutrino Number Density}\label{sec:DofsVanishing}

We now perform a similar analysis with the initial condition that $\eta^N(z_0) \simeq \overline{\eta}^N(z_0) \simeq 0$ for $z=z_0 \ll 1$. Hence $\Delta \simeq -\mathds{1}$ and $\delta\simeq 0$. In such scenarios, it can be seen that the dominant contribution to the TEs comes from the inverse decay of leptons into heavy neutrinos, which drive the system into thermal equilibrium. Since the effects of the variations in the {\em dofs} are of significance when the bath has an appreciable abundance, the deviation in the BAU derived from variations in the {\em dofs} is directly related to how quickly the system can generate a considerable number density for the heavy neutrino species. 

To make this last effect explicit, we first consider the initial evolution of the heavy neutrino number density. In addition, we may make some pertinent simplifications. Firstly, since the bath begins with $\eta^N \simeq 0$, we may neglect the contributions to the TEs from variations in $\eta^N_{\rm eq}(z)$, and $h_{\rm eff}(z)$. Similarly, the fact that the heavy neutrino number density is so far away from equilibrium means that we only need to consider inverse decays in the TEs since these will occur significantly more rapidly than the decay process. Finally, since the commutator term of the TEs does not increase or decrease the particle number, we may also neglect these terms. With these simplifications in mind, the TEs may then be written in the compact form
\begin{equation}
    \frac{d \underline{\eta}^N}{d \ln z} = \frac{1}{H n^\gamma} \Re{\gamma^N_{L\Phi}}, \qquad \frac{d \delta\eta^N}{d \ln z} = \frac{2i}{H n^\gamma} \Im{\delta\gamma}\,.
\end{equation}
Integrating the former of these expressions, we see that for $z\ll 1$,
\begin{align}
    \underline{\eta}^N(z)\, &=\, \int_{0}^z \, \frac{1}{H(z^\prime) n^\gamma(z^\prime)} \Re{\gamma^N_{L\Phi}(z^\prime)} \, \frac{dz^\prime}{z^\prime}\, \simeq\, \frac{1}{2\zeta(3)} \frac{\Re{\Gamma_T}}{H(z=1)}\int_{0}^z \, z^\prime K_1(z^\prime) \, dz^\prime \nonumber\\
    &\simeq\, \frac{1}{\zeta(3)} \frac{\Re{\Gamma_T}}{H(z=1)} \frac{z^3}{6}\;.
\end{align}
For low values of $z$, the value of $\eta^N_{\rm eq} \simeq \mathds{1}/\zeta(3)$ is close to constant. We define the system as close to equilibrium by the condition
\begin{equation}
    \Tr{\underline{\eta}^N(z_{\rm eq})}\, \simeq\, \Tr{\eta^N_{\rm eq}(z_{\rm eq})}\, .
\end{equation}
Using this last expression and assuming a democratic flavour structure $a=b=c = |\boldsymbol{h}^\nu|$ and constant {\em dofs}, i.e.~$g_{\rm eff}(z=1) \simeq 105$, we may estimate the equilibrium value, $z_{\rm eq}$, to be
\begin{equation}
  \label{eq:zeq}
    z_{\rm eq}\, \approx\, \Bigg(6 \frac{H(z=1)}{\textrm{Tr}\, \Re{\Gamma_T}}\Bigg)^{1/3} \simeq\, 13.7 \times \lrb{\frac{m_N}{M_{\rm Pl}}}^{1/3} |\boldsymbol{h}^\nu|^{-2/3}\, .
\end{equation}
From~\eqref{eq:zeq}, we observe that the system will reach thermal equilibrium early in the evolution when the Yukawa couplings are large and the heavy neutrino mass scale is low. This result is to be expected since these two conditions give the most preferable conditions for the inverse decay of SM leptons and Higgs bosons into heavy neutrinos. In the TRL model, the neutrino-Yukawa couplings have been estimated to be of order $10^{-4}$~\cite{daSilva:2022mrx}, and therefore it is expected that $1 \, \TeV$ heavy neutrinos will be close to equilibrium for $z_{\rm eq} \sim 10^{-1}$. Since the approach to equilibrium occurs faster for lower-scale heavy neutrinos, we may take this value of $z_{\rm eq}$ as an upper bound. Consequently, we may conclude that in strong washout regimes such as the one we present, the variations in the {\em dofs} can still be relevant, even when the neutrino number density begins far from equilibrium. However, their effect may only be significant once an appreciable number density of heavy neutrinos has been generated through inverse decays.

\subsection{Attractor Trajectories of the Transport Equations}

\begin{figure}[t]
    \centering
    \includegraphics[width=0.9\linewidth]{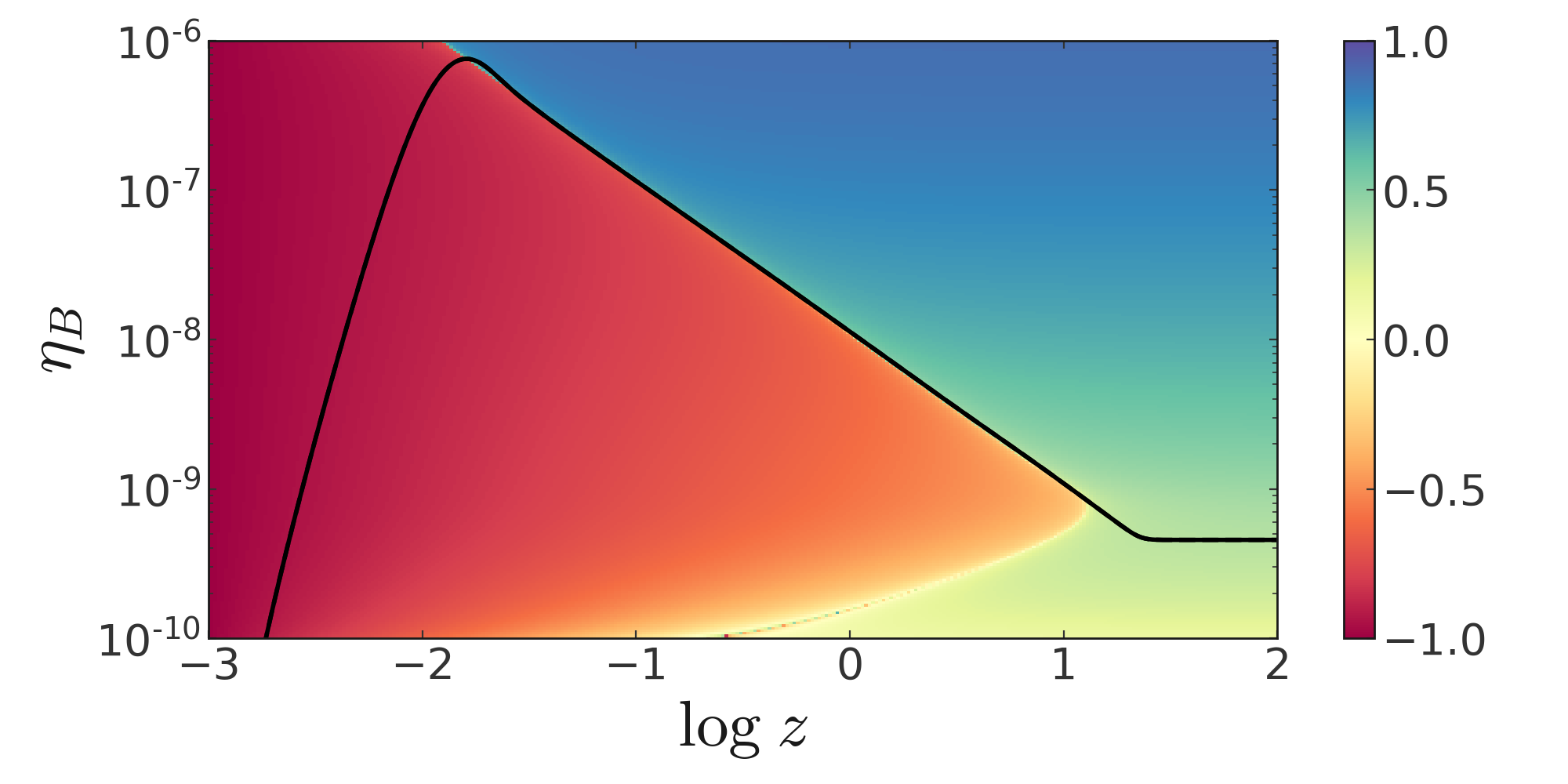}
    \caption{The attractor properties of the numerical solution to the TEs. Red shades indicate regions where the baryon asymmetry is increasing, and blue shades indicate regions where the baryon asymmetry is decreasing. Yellow shades indicate regions where the changes to $\eta_B$ are small.}
    \label{fig:Attractor}
\end{figure}

From the discussion in Section~\ref{sec:DofsVanishing}, the rapid approach to equilibrium of the heavy neutrino number density poses the question as to whether the TEs we utilise exhibit strong attractor properties. To address this question, we first consider the unusual initial condition, $\underline{\eta}^N(z_0) = 2\eta^N_{\rm eq} \mathds{1}$. As opposed to the scenario given in Section~\ref{sec:DofsVanishing}, the decay of heavy neutrinos to SM leptons and Higgs bosons becomes now the dominant channel. Moreover, this over-abundance of heavy neutrinos also enhances the dynamics from the {\em dofs}. Taking these two sources into account for $z\ll 1$, the heavy-neutrino transport equation may be approximated as follows:
\begin{equation}
   \label{eq:etaNattr}
    \frac{d \underline{\eta}^N}{d \ln z}\: \simeq\: - \frac{1}{H n^\gamma} \Re{\gamma^N_{L\Phi}}\, +\, 6\eta^N_{\rm eq} (1 - \delta_h)\ \mathds{1}\, .
\end{equation}
In order to assess whether or not the variations in the {\em dofs} are significant, we must compare the relative size of the two contributing terms that occur on the RHS of~\eqref{eq:etaNattr}. In this way, we deduce the condition
\begin{equation}
    \frac{|\boldsymbol{h}^\nu|^2\, m_N \, z^2}{8\pi H(z=1)}\,  \frac{K_1(z)}{K_2(z)}\: \gg\:  3(1 - \delta_h) \,.
\end{equation}
Taking a trace and making some appropriate approximations~\cite{Buchmuller:2004nz}, we see that the decay of heavy neutrinos will dominate the evolution when
\begin{equation}
     z^4 |\boldsymbol{h}^\nu|^2 \lrb{\frac{M_{\rm Pl}}{m_N}}\times 10^{-3} \gg  3(1 - \delta_h) \, .
\end{equation}
As may be inferred from Figure~\ref{fig:dH}, the value of the RHS of this last inequality typically lies in the range $10^{-4} - 10^{-3}$ for low values of $z$. Again assuming a Yukawa coupling scale of $|\boldsymbol{h}^\nu| \sim 10^{-4}$ and a heavy neutrino mass scale $m_N \simeq 1\,\TeV$, we see that the LHS is dominant for~${z>10^{-2}}$. Consequently, we may conclude that the effect of the {\em dofs} is drastically enhanced when the neutrino number density is close to equilibrium. Otherwise, the asymmetry between the rate of decays and inverse decays is the dominant effect, pushing the bath toward the equilibrium. 

By a similar argument to that given in Section~\ref{sec:DofsVanishing}, we may estimate again that $z_{\rm eq} \sim 10^{-1}$, but approaching from above rather than below. This indicates that the TEs we use may be highly attractive. Since we are now convinced that the variations in the {\em dofs} are only relevant close to chemical equilibrium, it is enough to verify that the system with no variations in the {\em dofs} exhibits such an attractor behaviour.

In all of our benchmark scenarios, we assume the Yukawa couplings to have a democratic flavour structure such that no SM lepton flavour is preferred, and flavour effects are minimised (i.e. $a=b=c$ in (\ref{eq:Yukawa})). In practice, this means that the model we adopt is similar to that of a single lepton generation. With this in mind, we study the attractor trajectories of a model with three singlet neutrino species and a single lepton flavour. In Figure~\ref{fig:Attractor}, we show the evolution of such a model, assuming a tri-resonant mass spectrum with $m_{N_1} = 1\,\TeV$ and $a=b=c=8.85\times 10^{-4}$. In addition, we do not stop the generation of baryon asymmetries at the sphaleron temperature and simply allow the freeze-out present in the figure to be achieved thermally. The colouring of the figure is determined through the quotient
\begin{equation}
    -\, \frac{d\,\eta_B/d \ln z }{\eta_B} \, ,
\end{equation}
normalised to lie in $[-1,1]$. The values indicate not only the strength of the attraction but also its direction, with bluer hues identifying a suppression of $\eta_B$ and redder hues indicating an enhancement. From Figure~\ref{fig:Attractor}, we see that the baryon asymmetry is driven upwards toward the attractor solution early in the evolution, $z \lesssim 10^{-2}$. Once the baryon asymmetry reaches the attractor solution, $10^{-2}\lesssim z \lesssim 10$, we see the most noticeable feature of this figure in the sharp transition from enhancement to suppression as we move across the attractor, highlighting the level to which this system of equations pushes the solution toward the attractive solution. Hence, we find that this set of TEs is highly attractive and exhibits independence of the initial conditions. Notice that Figure~\ref{fig:Attractor} also identifies when thermal freeze-out occurs at $z \sim 10$. Displayed through yellow shades, this region implies that $|\eta_B^\prime| \ll |\eta_B|$, and hence there is little change in the value of $\eta_B$ past this point in the evolution.

Finally, we identify that due to the inverse proportionality between $z_{\rm eq}$ and the Yukawa coupling scale, the rapid approach to equilibrium of the heavy-neutrino number density only applies to strong washout models of leptogenesis. In the case of weak washout models, such as the freeze-in models studied in \cite{Drewes:2022kap, Hernandez:2022ivz}, the time taken to reach equilibrium is far longer. In~such models, the out-of-equilibrium Sakharov condition relies on some unaccounted for pre-existing dynamics that set the initial heavy-neutrino number density to vanish in an {\it adhoc} manner at a convenient moment of the cosmic time during the evolution of the early Universe. Therefore, due to the slow approach to equilibrium, the contributions from the relativistic {\em dofs} may be overshadowed, with the generated BAU showing little deviations away from a model with constant {\em dofs}.

\subsection{Impact on the Baryon Asymmetry}\label{sec:DOfsBAU}

\begin{figure}[t]
    \begin{subfigure}{0.49\linewidth}
        \centering
        \includegraphics[width=\linewidth]{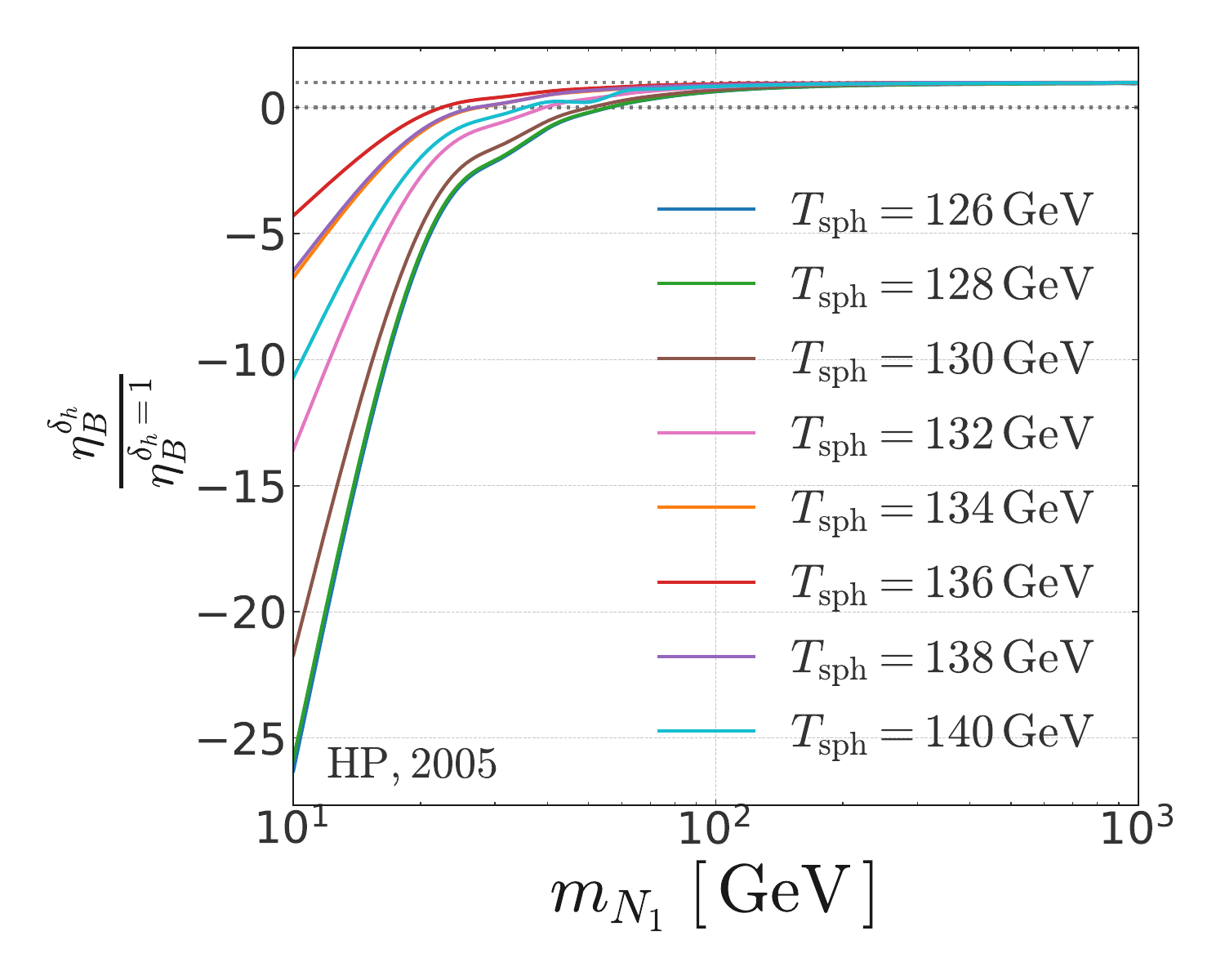}
        \caption{\empty}
    \end{subfigure}
    \begin{subfigure}{0.49\linewidth}
        \centering
        \includegraphics[width=\linewidth]{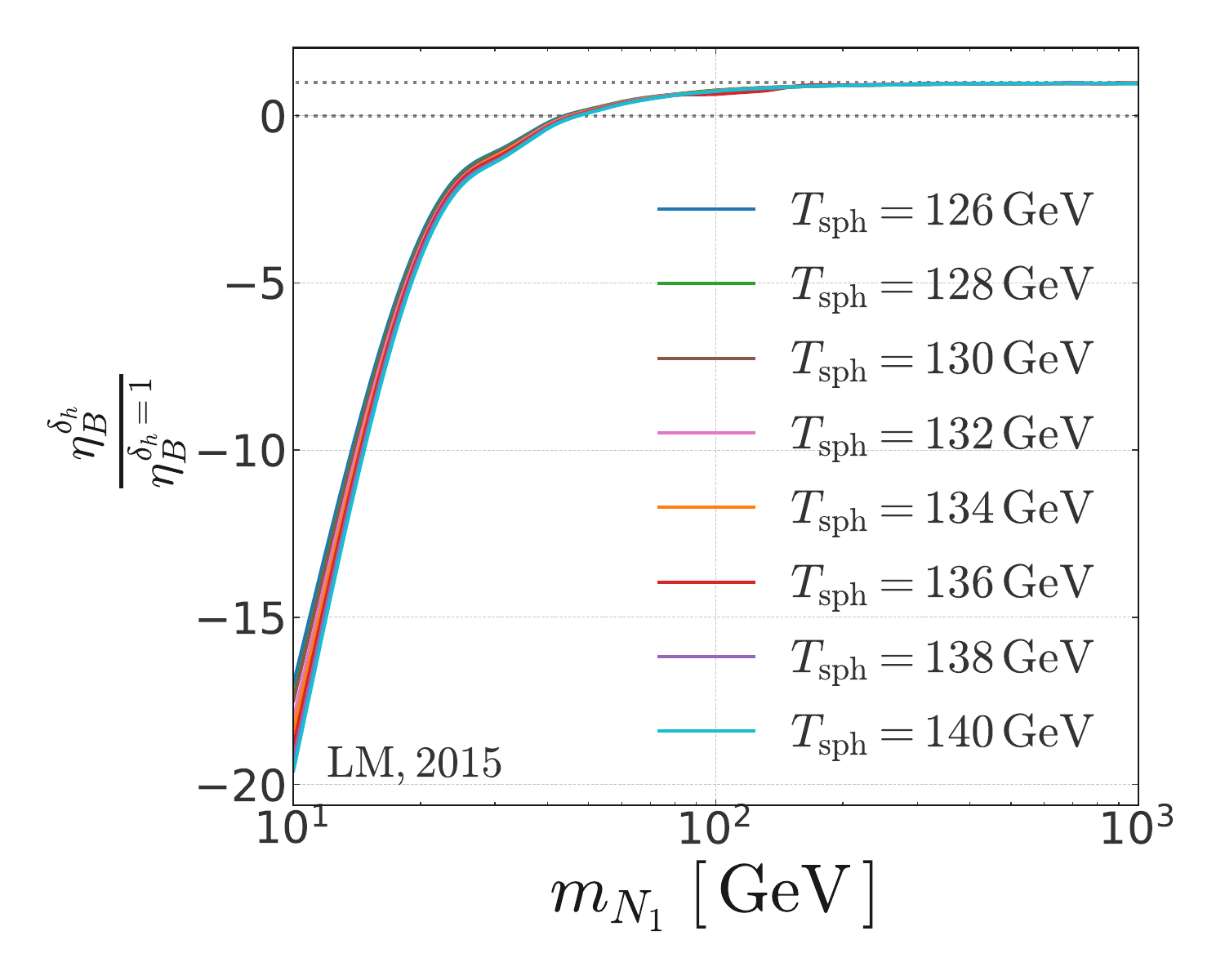}
        \caption{\empty}
    \end{subfigure}
    \caption{The ratio between the baryon asymmetry generated in a model with variations in the {\em dofs}, $\eta_B^{\delta_h}$ and the baryon asymmetry generated in a model where the {\em dofs} are constant $(h_{\rm eff} = 105)$, $\eta_B^{\delta_h=1}$. Different colours represent models where the sphaleron temperature takes different values. Two horizontal dotted lines are included to show unity and zero. The left panel is produced using data from Hindmarsh and Philipsen~\cite{Hindmarsh:2005ix} (HP, 2005), whereas the right panel utilises the latest data from Laine and Meyer~\cite{Laine:2015kra} (LM, 2015).}
    \label{fig:dofs}
\end{figure}
In previous sections, we have shown how the variations in the {\em dofs} may be significant in many leptogenesis scenarios. In this subsection, we will now discuss how these effects can impact the predictions for the BAU, including the interplay of sphaleron interactions that enter through the sphaleron freeze-out temperature~$T_{\rm sph}$. As previously noted, the largest contribution to the TEs from the variations in the {\em dofs} comes from an addition out-of-equilibrium term. This term washes out the number density through the freeze-out of the {\em dofs}. As a consequence, in regions where the variations in the {\em dofs} are dominant, we expect the neutrino number density to be under-abundant when compared to the equilibrium. Furthermore, since the transport equation for the lepton asymmetry is directly dependent on~$\Delta$, the lepton asymmetry follows a similar evolution. This leads to the less appealing feature that the generated baryon asymmetry turns suddenly negative, i.e.~$\eta_B < 0$, thus prompting us to resort to a contrived re-tuning of the CP phases of the model, often depending on the exact values of $T_{\rm sph}$ and $m_N$ considered. 

As was shown in Section~\ref{sec:DofsEqu}, models with a Majorana mass scale below $100\:\GeV$ may be highly sensitive to variations in the relativistic {\em dofs}. Figure~\ref{fig:dofs} displays how the generated BAU is altered by including the variations in the relativistic {\em dofs}. For this figure, the Yukawa couplings at each mass scale are taken such that the baryon asymmetry is reproduced by a system with no variations in the relativistic {\em dofs}, with democratic flavour structure assumed $(a=b=c=|\boldsymbol{h}^\nu_{ij}|)$. These Yuakwa couplings are then utilised in a system with the variations in the {\em dofs} included to find the generated BAU. Figure~\ref{fig:dofs} then shows the ratio between the two values of $\eta_B$. For clarity, grey dotted lines are shown on the figure at zero and unity. Finally, in order to better gauge the significance of the sphaleron effects, we also vary the sphaleron temperature widely around the typical value $T_{\rm sph} = 132\:\GeV$.

As can be seen in Figure~\ref{fig:dofs}, when the Majorana mass scale is above $100\:\GeV$, the generated BAU does not vary significantly from the constant {\em dofs} assumption. This occurs for two reasons. Firstly, as was discussed in Section~\ref{sec:DofsEqu}, the lack of deviation occurs at these mass scales because the changes to $\eta^N_{\rm eq}$ become the dominant out-of-equilibrium effect, and we expect independence from the variations in the relativistic {\em dofs}. Secondly, once the heavy neutrinos are out of equilibrium, the inverse decay of SM leptons and Higgs bosons will begin to drive the neutrino number density back towards equilibrium. This latter process weakens the influence of the initial dynamics from the {\em dofs}. Therefore, the inclusion of the variations in the relativistic {\em dofs} in models with $m_N > 100\:\GeV$ effectively generates new initial conditions for the TEs. For comparison, we observe that for $m_N<100\:\GeV$, large deviations are apparent. In a model with such low Majorana mass scales, the washout effect from the relativistic {\em dofs} is more significant than the change in the equilibrium number density, producing a number density below the equilibrium value. Moreover, early in the evolution, the interactions with the SM leptons and Higgs bosons do not have the necessary time to restore the number density to its equilibrium value. Consequently, we may find that the baryon asymmetry at this scale drops below zero.

Figure~\ref{fig:dofs} also indicates that regardless of the sphaleron critical temperature, the observed BAU drops below zero and takes on greater negative values as we approach lower mass scales. However, we point out that due to the complex interplay of the two sources of out-of-equilibrium dynamics, the deviations observed depend non-trivially on the sphaleron temperature. Due to the sign change in the generated BAU in low-scale mass regions, it is necessary to invert the CP phase in the Yukawa sector. In the $\mathbb{Z}_6$ model, this is easily achieved by the replacement $\omega \to \omega^*$ in the tree-level Yukawa couplings, which exchanges $\boldsymbol{h}_+^\nu \leftrightarrow \boldsymbol{h}_-^\nu$.

\begin{figure}[t]
    \centering
    \begin{subfigure}{0.49\linewidth}
        \includegraphics[width=\linewidth]{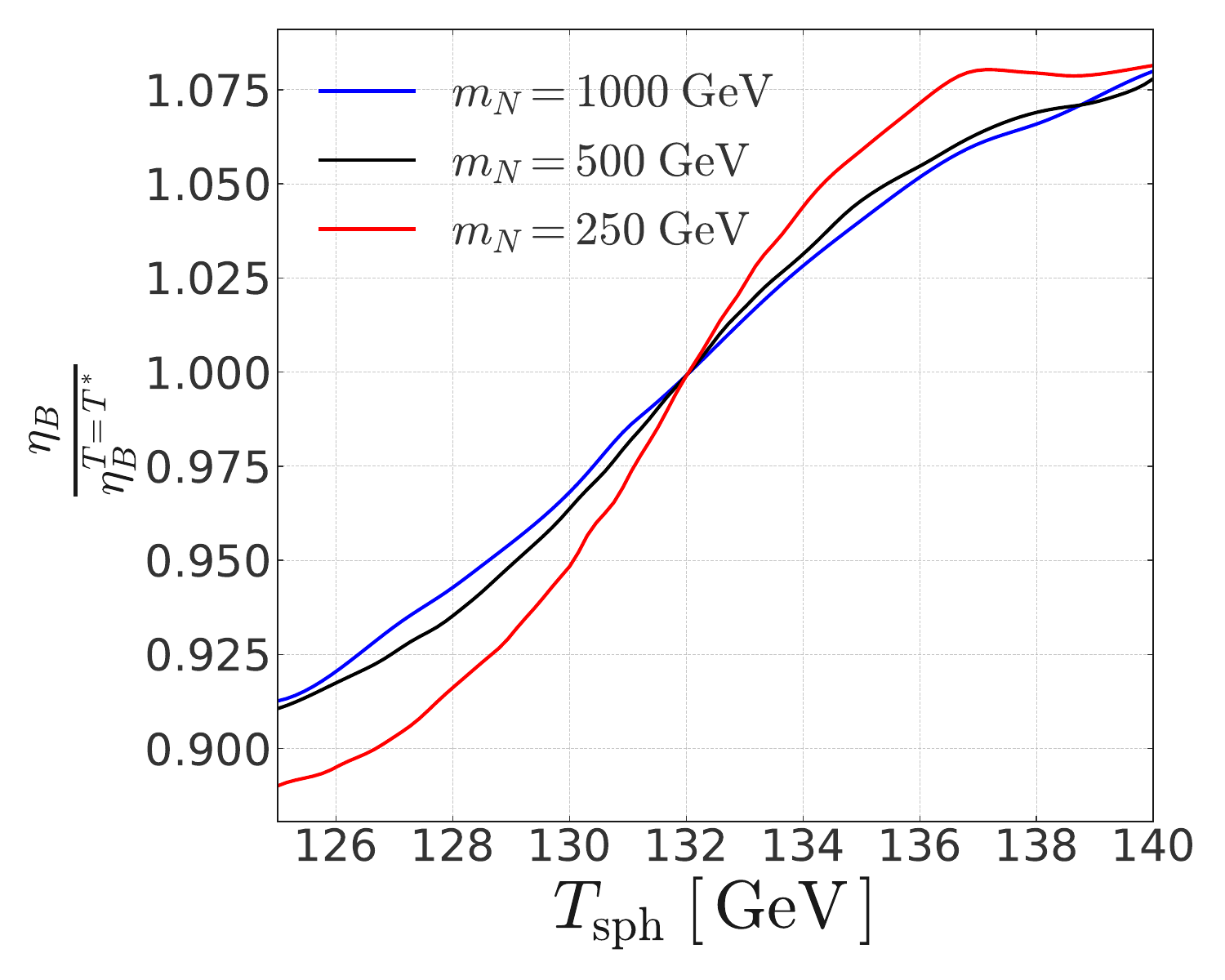}
        \caption{\empty}
        \label{fig:TempHiM}
    \end{subfigure}
    \begin{subfigure}{0.49\linewidth}
        \includegraphics[width=\linewidth]{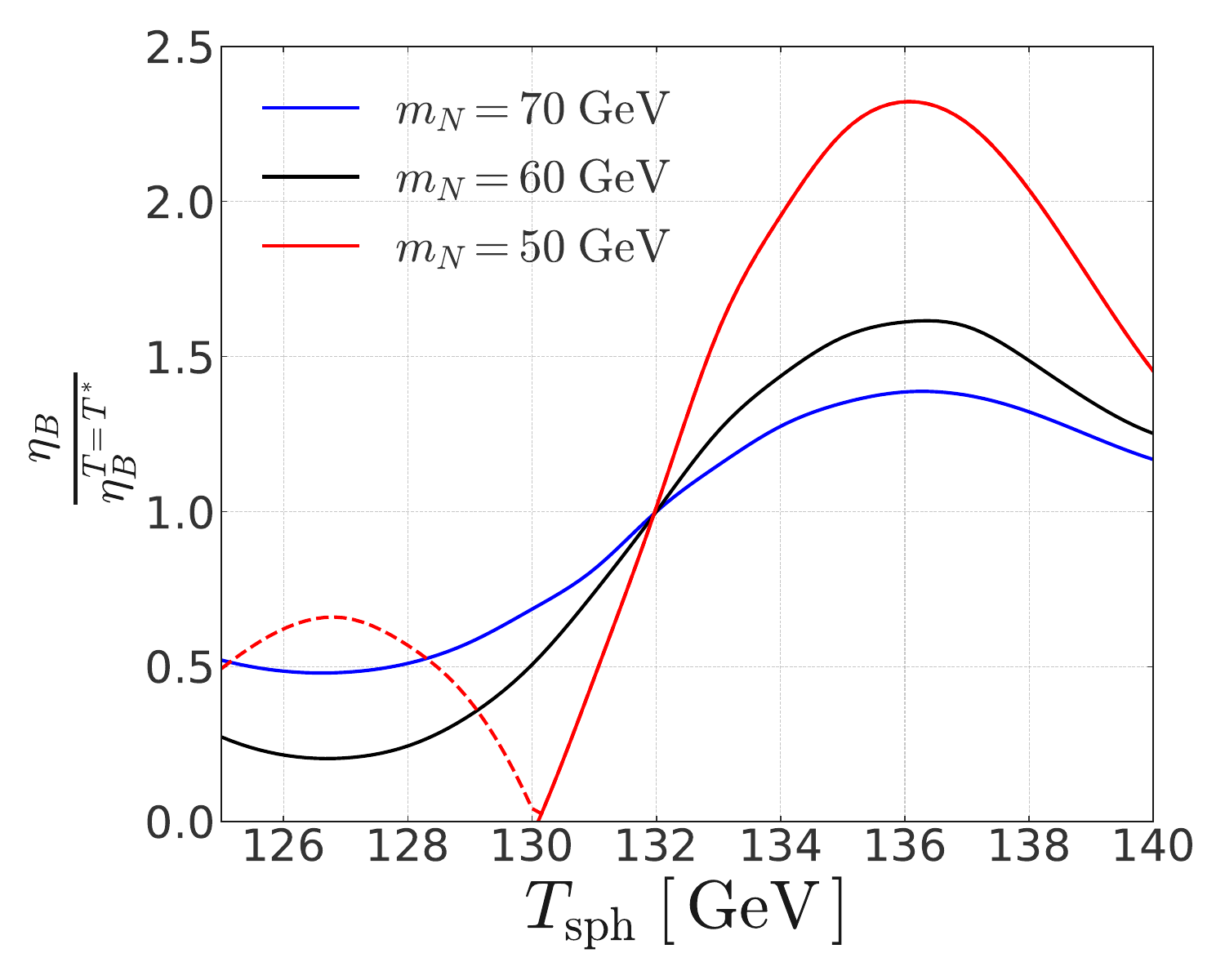}
        \caption{\empty}
        \label{fig:Temp50GeV}
    \end{subfigure}
    \caption{The dependence of the generated BAU on variations to the sphaleron temperature, $T_{\rm sph}$ when the \textit{dofs} are generated from~\cite{Hindmarsh:2005ix}. These figures show the ratio between the generated baryon asymmetry at a given sphaleron temperature, $\eta_B$, and the generated baryon asymmetry when the sphaleron critical temperature is taken to be $T_{\rm sph} = T^* = 132\:\GeV$, $\eta_B^{T = T^*}$. The left panel shows this ratio for heavy neutrino masses above the electroweak scale, and the right panel shows the ratio for low-scale models below the electroweak scale.}
    \label{fig:Temps}
\end{figure}

As already alluded to, depending on the parameterization for the {\em dofs}, there may be a complex interplay between the competing effects from the changes in the relativistic {\em dofs}, $h_{\rm eff}$, and changes in the equilibrium $\eta^N_{\rm eq}$. Consequently, there may be a complicated dependence of the generated BAU on the sphaleron temperature. Figure~\ref{fig:Temps} demonstrates this relationship when we use the parameterisation given in~\cite{Hindmarsh:2005ix}. To produce this figure, the Yukawa coupling necessary to reproduce the observed BAU at $T_{\rm sph} = 132\:\GeV$ was found for each mass scale considered. This Yukawa coupling was then used to estimate the BAU generated in models with different values of $T_{\rm sph}\in \lrsb{125,140}\, \GeV$, and the ratio between the two values of $\eta_B$ plotted.

Figure~\ref{fig:TempHiM} shows the dependence of $\eta_B$ on $T_{\rm sph}$ for high-scale Majorana masses. As expected, models at this scale show little dependence on the sphaleron critical temperature, with most scenarios showing less than $\sim 10\%$ deviation away from the value of $\eta_B$ with $T_{\rm sph}= 132\:\GeV$. However, at low mass scales, shown in Figure~\ref{fig:Temp50GeV}, the generated BAU is significantly more sensitive to changes in the sphaleron critical temperature. Again, at these scales, the relevant contribution from the variations in the {\em dofs} produces a non-trivial relationship between the observed BAU and the sphaleron critical temperature. Here, we see that the observed deviations are significantly larger, up to a factor of $2.3$ for $m_N = 50\:\GeV$. In line with previous results, we also see that for lower-scale heavy neutrino masses, larger deviations are present, and in some cases, a negative value for the baryon asymmetry may be generated (indicated through the dashed line), which again may be rectified through the inversion of the CP phases.

\begin{figure}
    \centering
    \includegraphics[width=0.6\linewidth]{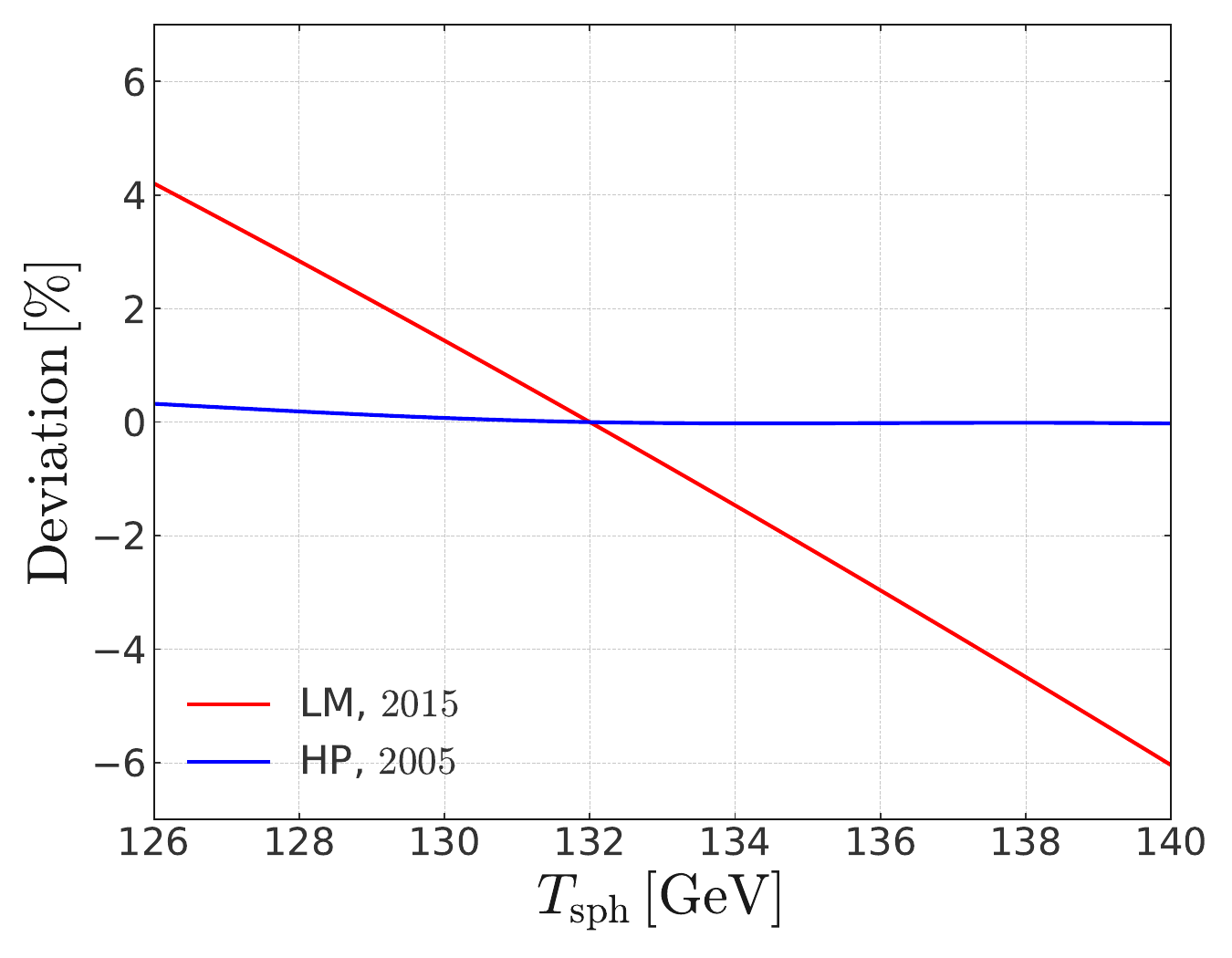}
    \caption{ The deviation in the washout factor, which converts the value of the BAU at $T_{\rm sph}$ to the observed value at $T_{\rm rec}$, as a function of $T_{\rm sph}$. The value of this factor at $T_{\rm sph}$ is taken as the reference value. The blue line is generated using the data from EOS C by Hindmarsh and Philipsen~\cite{Hindmarsh:2005ix} (HP, 2005), and the red line is made using the latest data from Laine and Meyer~\cite{Laine:2015kra} (LM, 2015).}
    \label{fig:SphaleronConversion}
\end{figure}

Another source of potential deviation arises from the dilution factor, which comes from the conversion of the baryon asymmetry value at the freeze-out value at $T_{\rm sph}$ to the observed value at $T_{\rm rec}$. As is shown in equation~(\ref{eq:dilution}), the conversion comes from a ratio between the number of degrees of freedom at $T_{\rm sph}$ and $T_{\rm rec}$, therefore a change in $T_{\rm sph}$ can alter the predictions for the baryon asymmetry. In Figure~\ref{fig:SphaleronConversion}, we show how this value of the conversion factor changes from the value at $T_{\rm sph} = 132\,\GeV$. As can be seen in the figure, when the parameterisation given in~\cite{Hindmarsh:2005ix} is used, the variations in the predictions for the BAU vary by less than $1\%$. Conversely, for the data extracted from~\cite{Laine:2015kra}, these variations may increase to up to $6\%$ at the more extreme values of the sphaleron critical temperature.

As a final point, it is worth re-iterating that the parameterisation of the variations in the relativistic {\em dofs} may be of significance. As was shown in~\cite{daSilva:2022mrx}, using a different data set for the {\em dofs} may affect the predictions for the BAU. In particular, when using~\cite{Gondolo:1990dk}, it was shown that the deviations had a lesser effect on the generated BAU, although notable features such as sudden parametric changes in the sign of $\eta_B$ are still present. In this work, we have made additional comparisons of the predicted Baryon asymmetry when using~\cite{Hindmarsh:2005ix}, which offers an improved data set through better modelling of the QCD quark-gluon plasma, or \cite{Laine:2015kra} which carefully considers distinct thermal regions, as well as a phenomenologically accurate Higgs mass. In general, these two data sets exhibit similar behaviour; however, the more recent data given in~\cite{Laine:2015kra} seems to produce more robust predictions for the BAU with less dependence on the assumed value of the sphaleron critical temperature, as is highlighted in figure~\ref{fig:dofs}, demonstrating the importance of accuracy in the determination of cosmological parameters.

\section{Critical Scenarios}\label{sec:Crit}
\setcounter{equation}{0}
In~\cite{Karamitros:2022oew}, the behaviour of unstable two-level quantum systems, famously called {\em qubits}, was discussed in detail. The dynamics of an unstable multi-level quantum system, also known as {\em qudit} in quantum information theory, is typically described by employing the Weisskopf-Wigner approximation (WW)~\cite{Weisskopf:1930au}. In the WW approximation, the states of a given quantum system evolve under a Schr\"{o}dinger-like equation, with effective Hamiltonian
\begin{equation}
    \textrm{H}_{\rm eff}\: =\: \textrm{E}\, -\, \frac{i}{2}\Gamma\,.
\end{equation}
Here, $\textrm{E}$ and $\Gamma$ are Hermitian matrices corresponding to the self-energy corrected energy matrix\- and the absorptive part of the self-energy correction to a multi-level particle system, respectively. Using this effective Hamiltonian, it may be shown that the density matrix of the multi-level system follows the evolution equation
\begin{equation}\label{eq:Liouville}
    \frac{d \rho}{dt}\: =\: -i \lrsb{\textrm{E}, \rho}\, -\, \frac{1}{2}\lrcb{\Gamma, \rho}\, .
\end{equation}
However, by taking the trace of this expression, it can be seen that this evolution equation loses probability as the states of the multi-level system decay to states not contained within the Hilbert space under consideration. As a result, the density matrix, which only contains the information pertinent to the states which have not yet decayed, is modified from the typical definition. This adjustment is done through the trace of the density matrix which we employ to define the {\em co-decaying} density matrix
\begin{equation}
    \hat{\rho}\, \equiv\, \frac{\rho}{\textrm{Tr} \, \rho} \, .
\end{equation}
The latter obeys the evolution equation
\begin{equation}
    \frac{d \hat{\rho}}{dt}\: =\: -i \lrsb{\textrm{E}, \hat{\rho}}\, -\, \frac{1}{2}\lrcb{\Gamma, \hat{\rho}}\, +\, \hat{\rho}\Tr{\hat{\rho}\Gamma} \, ,
\end{equation}
for which it may be seen that the trace of the evolution equation vanishes provided $\textrm{Tr} \, \hat{\rho} = 1$,
\begin{equation}
    \frac{d \, {\rm Tr} \hat{\rho}}{dt} = 0 \, ,
\end{equation}
implying conservation of the total probability of the co-decaying density matrix. 

In the case of a two-level system, the matrices we consider may be expanded on the Pauli basis through four real-valued coefficients
\begin{equation}
   \label{eq:CritScenMats}
    \hat{\rho} = \frac{1}{2}b_\mu \overline{\sigma}^\mu \, , \quad \textrm{E} = E_\mu \sigma^\mu \, , \quad \Gamma = \Gamma_\mu \sigma^\mu \, ,
\end{equation}
where $b^\mu = \lrb{1,\boldsymbol{b}}$, $\sigma^\mu = \lrb{\mathds{1}, \boldsymbol{\sigma}}$, $\overline{\sigma}^\mu = \lrb{\mathds{1}, -\boldsymbol{\sigma}}$, $E^\mu = \lrb{E^0, \boldsymbol{E}}$, and $\Gamma^\mu = \lrb{\Gamma^0, \boldsymbol{\Gamma}}$. Note that in this section, boldface identifies vector-like quantities rather than objects with flavour structure. From these definitions, an equation which describes the motion of the Bloch vector, $\boldsymbol{b}$, may be derived
\begin{equation}
    \frac{d \boldsymbol{b}}{d\tau}\: =\: -\frac{1}{r} \boldsymbol{e}\times \boldsymbol{b}\, +\, \boldsymbol{\gamma}\, -\, \lrb{\boldsymbol{\gamma}\cdot \boldsymbol{b}}\boldsymbol{b}\, .
\end{equation}
In this expression, two dimensionless parameters are introduced
\begin{equation}
    r = \frac{|\boldsymbol{\Gamma}|}{2|\boldsymbol{E}|},\quad \tau = |\boldsymbol{\Gamma}|t \, ,
\end{equation}
as well as the unit vectors
\begin{equation}
    \boldsymbol{e} = \frac{\boldsymbol{E}}{|\boldsymbol{E}|}\, , \quad \boldsymbol{\gamma} = \frac{\boldsymbol{\Gamma}}{|\boldsymbol{\Gamma}|} \, .
\end{equation}
In~\cite{Karamitros:2022oew}, it was shown explicitly that in the majority of cases, the system will evolve into a pure state with a stationary co-decaying Bloch vector, $\boldsymbol{b}_\star = \boldsymbol{b}(\tau \to \infty)$, corresponding to the state with the longest lifetime. However, 
under special circumstances for which $\boldsymbol{e}\cdot\boldsymbol{\gamma} = 0$ and $r<1$, it may be seen that there is no preferred state of the qubit system since the lifetime for both states is identical. As a result, the co-decaying Bloch vector never reaches an asymptotic limit and oscillates between states indefinitely. Classes of models which satisfy the two conditions $\boldsymbol{e}\cdot\boldsymbol{\gamma} = 0$ and $r<1$ are referred to as {\em critical scenarios} and exhibit a number of interesting properties. In particular, one finds oscillations between coherent and decoherent states as well as the anomalous oscillations of the Bloch vector sweeping out {\em unequal} areas in equal time, in contrast with the typical Rabi oscillations for stable particle systems. It may be the case that the energy and decay vectors are close to orthogonal and $r<1$, and so it may still be possible to observe some critical phenomena such as coherence-decoherence oscillations. However, in such cases, the system is expected to eventually relax into its longest-lived state.

\subsection{Critical Scenarios in the Thermal Plasma}

We now investigate whether similar phenomena can be present within cosmological settings. Comparing the TEs given in equation~(\ref{eq:TEslnz}) with the evolution equation given in equation~(\ref{eq:Liouville}), we notice a significant overlap between the terms present. However, in the TEs, the matrix we are interested in would be that of the departure-from-equilibrium matrices, $(\Delta, \delta)$ [c.f.~\eqref{eq:DFEMats}], rather than the density matrix. Therefore, any phenomena we observe occur about the equilibrium number density. 

In order for us to get a better insight of the relevant dynamics, it would be preferable to work with a single TE rather than two interdependent equations. Since we expect that interesting dynamics will occur about the equilibrium, we would like to retain the use of the matrices $\Delta$ and $\delta$. Consequently, we give a definition close to that of the distribution functions, $f^N$ and~$\overline{f}^N$, which are linear combinations of the CP-even and CP-odd parts. To this end, we define
\begin{equation}
    D_{\pm}\: \equiv\: \Delta\, \pm\, \frac{1}{2}\delta\;,
\end{equation}
in terms of which the following TE can be derived:
\begin{multline}
    \frac{d D_{\pm}}{d \ln z} = \frac{1}{H(z)} \lrb{\mp i \lrsb{\mathcal{E}^N, D_{\pm}} - \frac{1}{2n^\gamma \eta^N_{\rm eq}} \lrcb{D_{\pm}, \Re{\gamma^N_{L\Phi}} \pm i \Im{\delta\gamma}}} \\
    + \lrb{3(1 - \delta_h) - \frac{d \ln \eta^N_{\rm eq}}{d \ln z}}\lrb{D_{\pm} + \mathds{1}} \, .
\end{multline}
For definiteness, we study the behaviour of the $D_+$ matrix. However, from the CP properties of $\Delta$ and $\delta$, we know that $D_+$ is related to $D_-$ through its transpose. In order to study the approach to equilibrium, we normalise the $D_+$ matrix by its trace
\begin{equation}
    \hat{D}_+\, =\, \frac{D_+}{\textrm{Tr} \, D_+} \, .
\end{equation}
Computing the TE for $\hat{D}_+$, we find the expression
\begin{multline}
    \frac{d \hat{D}_+}{d \ln z} = \frac{1}{H(z)} \left(- i \lrsb{\mathcal{E}^N, \hat{D}_{+}} - \frac{1}{2n^\gamma \eta^N_{\rm eq}} \lrcb{\hat{D}_{+}, \Re{\gamma^N_{L\Phi}} + i \Im{\delta\gamma}}\right.\\ 
    \left.+ \hat{D}_+ \Tr{\hat{D}_+ \lrb{\Re{\gamma^N_{L\Phi}} + i \Im{\delta\gamma}}}\right) + \lrb{3(1 - \delta_h) - \frac{d \ln \eta^N_{\rm eq}}{d \ln z}} \frac{1}{\textrm{Tr} \, D_+}\lrb{\mathds{1}- \hat{D}_+ N} \, ,
\end{multline}
where $N$ is the number of heavy neutrino generations. From this TE, we can see that when the heavy neutrino number density is close to its equilibrium value, only the final term contributes in a significant way since the trace in the denominator enhances its contribution. As a result, the normalised matrix $\hat{D}_+$ must be close to a full mixed state
\begin{equation}
    \hat{D}_+\, \simeq\, \frac{1}{N} \mathds{1} \, .
\end{equation}
Therefore, it would seem that the inclusion of terms which pull the number density away from equilibrium also drive the system toward a mixed state with no flavour preferences.

However, we notice that the remaining terms are similar to the evolution equation for a co-decaying Bloch vector. Therefore, we expect that in thermo-static backgrounds, it may be possible to observe some critical phenomena. Since we are now considering scenarios of constant temperature, the dimensionless parameter $z$ is no longer fit for use. Consequently, we parameterise the evolution through the cosmic time. Starting from equation~(\ref{eq:TEsTime}), we can find the evolution equation for the departure from equilibrium matrices as functions of time
\begin{equation}
    \frac{d D_+}{dt} = - i \lrsb{\mathcal{E}^N, D_{+}} - \frac{1}{2n^\gamma \eta^N_{\rm eq}} \lrcb{D_{+}, \Re{\gamma^N_{L\Phi}} + i \Im{\delta\gamma}} \, .
\end{equation}
This expression can clearly be mapped onto Equation~(\ref{eq:Liouville}), with the substitutions
\begin{equation}
    \rho\, \to\, \Delta\,, \qquad \textrm{E}\, \to\, \mathcal{E}^N\,, \qquad \Gamma\, \to\, \frac{1}{n^\gamma \eta^N_{\rm eq}}\lrb{\Re{\gamma^N_{L\Phi}} + i \Im{\delta\gamma}} \, .
\end{equation}
After these replacements, we can expect the Bloch vector for the normalised matrix $\hat{D}_+$ to act analogously to the co-decaying Bloch vector of an unstable qubit. By following a similar procedure to that of the qubit system and taking definitions in line with those given in~\ref{eq:CritScenMats}), 
\begin{subequations}
    \begin{equation}
    \mathcal{E}^N_\mu \sigma^\mu\, =\, \mathcal{E}^N, \quad E_\mu \sigma^\mu\, =\, \boldsymbol{m}_M \, ,
\end{equation}
\begin{equation}
    \gamma_\mu \sigma^\mu\, =\, \Re{\gamma^N_{L\Phi}} + i \Im{\delta\Gamma}, \quad \Gamma_\mu \sigma^\mu = \Re{\Gamma_T} + i \Im{\delta\Gamma} \, ,
\end{equation}
\end{subequations}
we find that the $r$ parameter for this system is given by
\begin{equation}
    r\, =\, \frac{|\boldsymbol{\gamma}|}{2n^\gamma \eta^N_{\rm eq}|\boldsymbol{\mathcal{E}}^N|}\, =\, \frac{|\boldsymbol{\Gamma}|}{2|\boldsymbol{E}|} \, .
\end{equation}
In addition, the $\tau$ parameter satisfies the transformation
\begin{equation}
    \tau\, =\, 
    \frac{K_1(z)}{K_2(z)}|\boldsymbol{\Gamma}| \, t \, .
\end{equation}

The $\mathbb{Z}_{2N}$ models under consideration, with the assumed resonant mass splitting, naturally give energy and decay vectors which are close to orthogonal. We consider a Yukawa matrix with entries given by
\begin{equation}
    \boldsymbol{h}^\nu_{i\alpha}\, =\, \ell_i\, \omega^{\alpha-1} \, ,
\end{equation}
where $\ell_i \in\mathbb{C}$ are complex parameters and $\omega = e^{i\psi}$ is simply a phase factor, with $\psi \in \mathbb{R}$. In~this case, the tree-level decay matrix is given by
\begin{equation}
   \label{eq:Gomega}
    \Gamma_{\alpha\beta}\, =\, \frac{m_N}{8\pi} \lrb{\sum_k |\ell_k|^2} \omega^{\beta-\alpha}\, . 
\end{equation}
It is not difficult to see from~\eqref{eq:Gomega} that the diagonal elements~$\Gamma_{\alpha\alpha}$ of the tree-level decay matrix are all identical. Therefore, with a structure like this, the only way to induce differences in the diagonal elements is through higher-order corrections, such as through the mixing of heavy-neutrino species. Consequently, we expect the coefficients of all $\textrm{SU}(N)$ diagonal generators of the decay matrix to vanish, and only the coefficients corresponding to off-diagonal $\textrm{SU}(N)$ generators will take non-zero values. For $N=2$, we usually have that the component 
$\Gamma_3$ of the four-vector $\Gamma_\mu$ is zero, but the other components $\Gamma_1$ and $\Gamma_2$ are non-zero [cf.~\eqref{eq:CritScenMats}]. Instead, the converse will hold true for the mass-matrix or energy components $E_{1,2,3}$, where $E_{1,2}$ are zero, but $E_3$ is not.
The latter is a simple consequence of the fact that 
the Majorana mass matrix is taken to be diagonal, and so all contributions to the energy vector~$\boldsymbol{E}$ appear as coefficients of the diagonal generators of the $\textrm{SU}(N)$ group. Thus, for a model with $N=2$ heavy Majorana neutrinos, this coefficient is proportional to the third Pauli matrix $\sigma^3$. 

It is important to note here that for models akin to the TRL $\mathbb{Z}_{2N}$ neutrino-Yukawa structure, the orthogonality of the vectors $\boldsymbol{E}$ and $\boldsymbol{\Gamma}$ is automatically satisfied by construction, i.e.~$\boldsymbol{E}\cdot \boldsymbol{\Gamma} = 0$. Moreover, computations of $r$ in these models yield values that are very close to~1. Therefore, the TRL model can serve as a prototypal example of a {\em non}-diagonalisable Jordan-form model, details of which are discussed in~\cite{Pilaftsis:1997dr, Karamitros:2022oew} and Appendix~\ref{App:Jordan}.

\begin{figure}
    \centering
    \begin{subfigure}{0.49\linewidth}
        \centering
        \includegraphics[width=\linewidth]{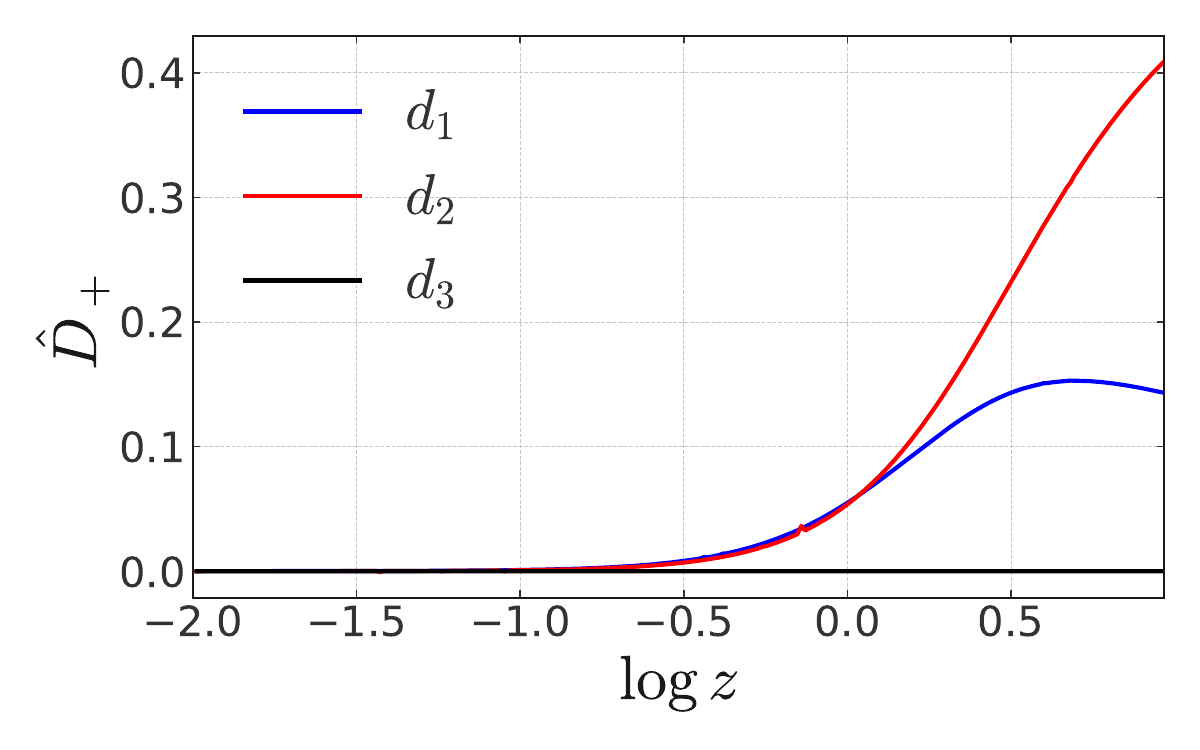}
        \caption{\empty}
        \label{fig:NonCS}
    \end{subfigure}
    \begin{subfigure}{0.49\linewidth}
        \centering
        \includegraphics[width=\linewidth]{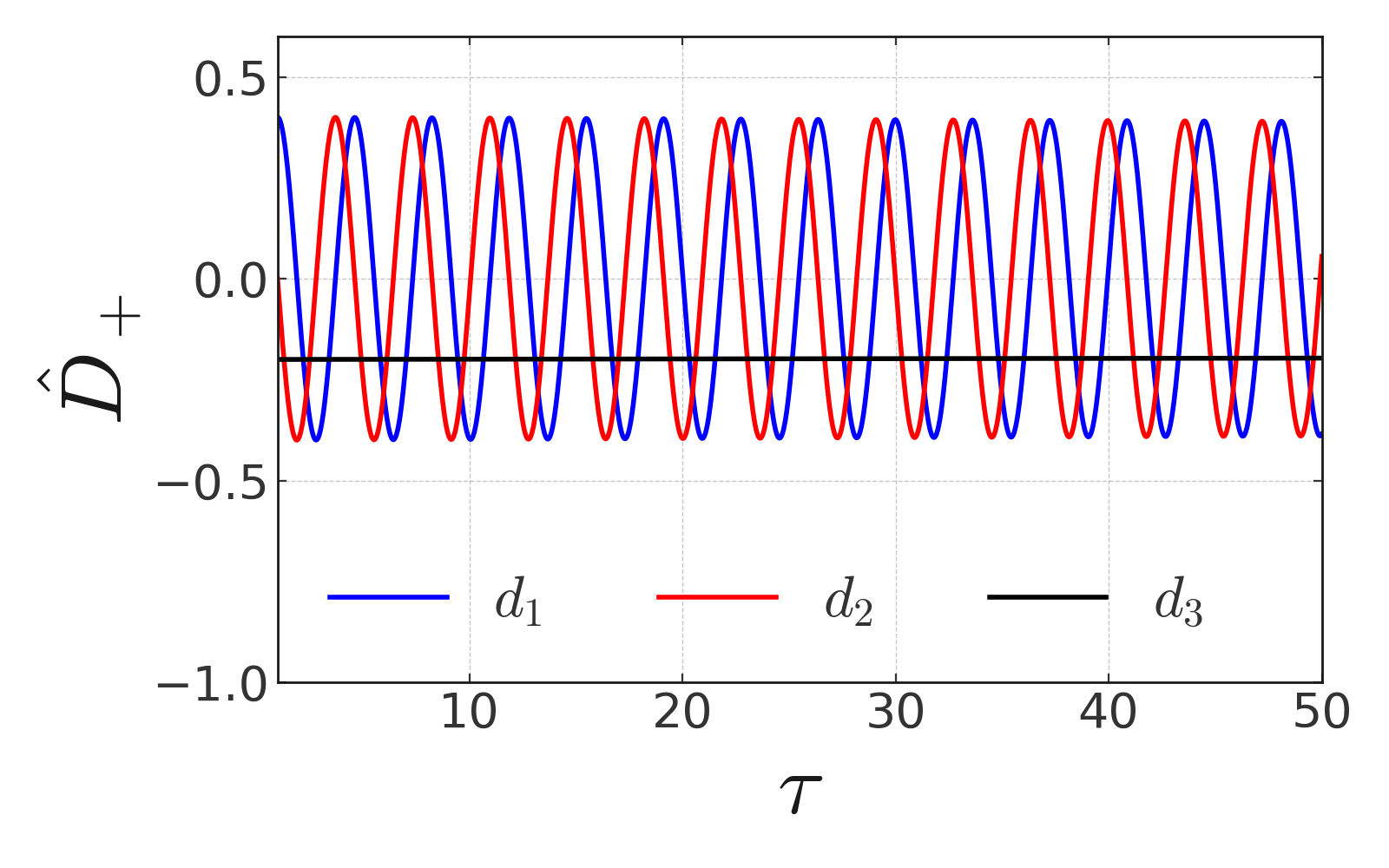}
        \caption{\empty}
        \label{fig:CS}
    \end{subfigure}
    \caption{The prevalence of critical phenomena in two different scenarios. The left panel shows the components of the co-equilibrium Bloch vector, $\hat{D}_+$, in a cosmological setting where out-of-equilibrium phenomena are present. The evolution is given in terms of the cosmological parameter $z$, since the cosmic time is not fixed through the expansion of the Universe.} The right panel shows the components of the co-equilibrium Bloch vector in a thermo-static Universe where the temperature is kept constant.
    \label{fig:Critical}
\end{figure}

In Figure~\ref{fig:Critical}, we display the dynamics of the $D_+$ matrix. We write the $\hat{D}_+$ matrix in its Bloch decomposition
\begin{equation}
    \hat{D}_+ = \frac{1}{2}\lrb{\mathds{1}+\boldsymbol{d}\cdot\boldsymbol{\sigma}} \, ,
\end{equation}
and display the elements of the {\em co-equilibrium} Bloch vector, $\boldsymbol{d}$. Figure~\ref{fig:NonCS} shows the evolution of the co-equilibrium Bloch vector in a standard cosmological setting where the Universe cools through its expansion. We use a model with two heavy neutrino species which have resonant mass splitting and Yukawa scales, $|\boldsymbol{h}^\nu|=4.5\times 10^{-4}$. Since the use of $\mathbb{Z}_4$ generators would lead to no lepton asymmetry, we use the $\mathbb{Z}_6$ generator as the phase in the Yukawa coupling. In Figure~\ref{fig:NonCS}, we observe that at the early phases of the evolution, the elements of the co-equilibrium Bloch vector lie close to zero, implying the number density matrix is fully mixed with no preference in the states. But later in the evolution, one may notice some dynamics arising from the oscillation and decay terms of the TEs. The latter results from the out-of-equilibrium dynamics which becomes less significant for the evolution since the decay of heavy neutrinos will occur quickly when compared with the expansion of the Universe. In this late stage, we see that the co-equilibrium Bloch vector no longer describes a fully mixed state but instead asymptotically approaches a non-zero state.

Figure~\ref{fig:CS} shows the dynamics in a thermo-static background. For this figure, we take the energy and decay vectors, $\boldsymbol{E}$ and $\boldsymbol{\Gamma}$, to be orthogonal to one another with $r=0.5$. The evolution is shown in terms of the dimensionless run parameter, $\tau$.  Here, the dynamics of critical scenarios are clearly present, in which oscillations between states take place indefinitely and the co-equilibrium Bloch vector never relaxes into an asymptotic state. Moreover, we see that the oscillations occur only along the directions $d_1$ and $d_2$, with the value in the final direction remaining constant, highlighting a defining property of critical scenarios in that the oscillations occur on a plane.

In the next section, we will give numerical estimates of the BAU predicted in the TRL model, including potential constraints from high- and low-energy observables of charged lepton-flavour and lepton-number violation. 

\section{Numerical Estimates}\label{sec:Results}
\setcounter{equation}{0}
\begin{figure}[t]
    \centering
    \begin{subfigure}{0.49\linewidth}
        \includegraphics[width=\linewidth]{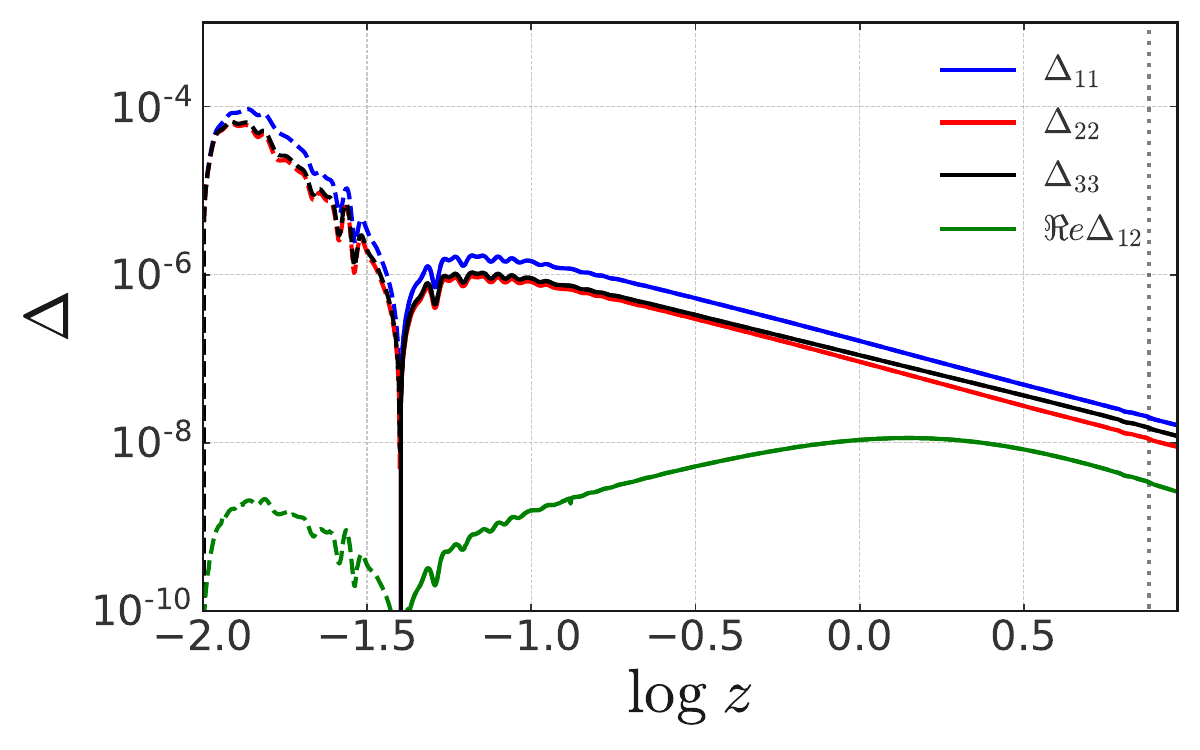}
        \caption{\empty}
        \label{fig:NeuEvolsTeV}
    \end{subfigure}
    \begin{subfigure}{0.49\linewidth}
        \includegraphics[width=\linewidth]{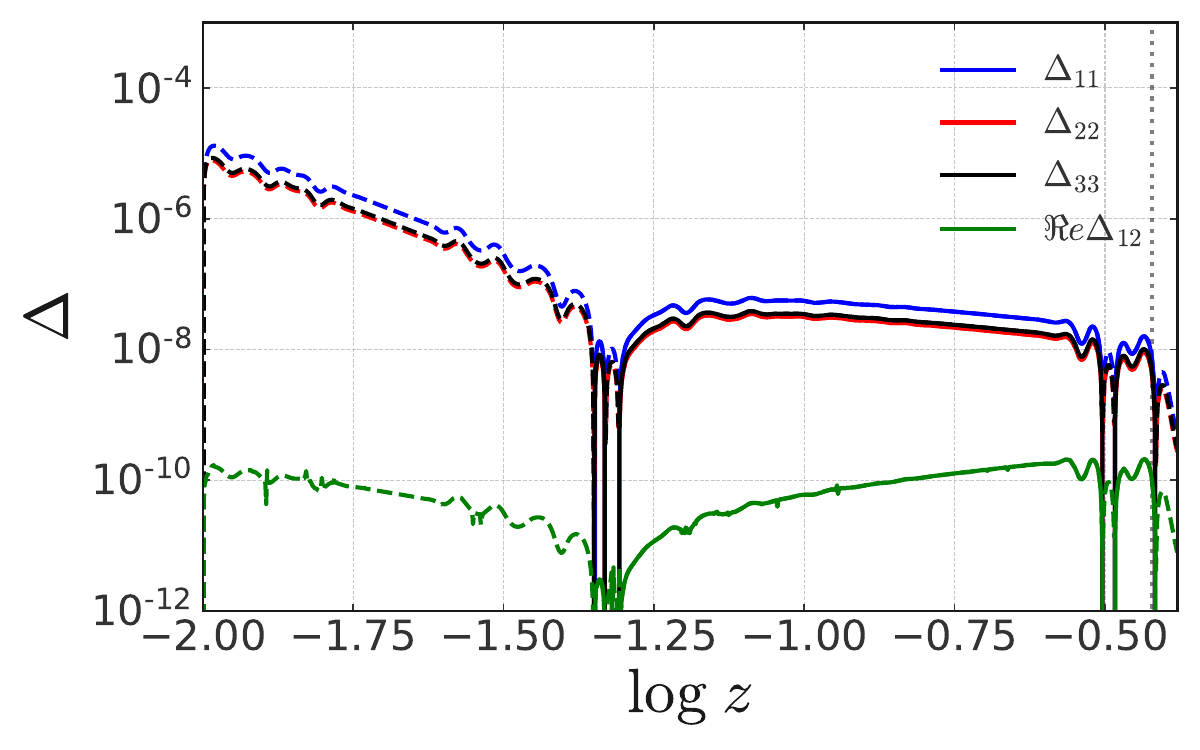}
        \caption{\empty}
        \label{fig:NeuEvols50}
    \end{subfigure}
    \caption{Numerical solutions for the heavy neutrino departure from equilibrium matrices at two points in the parameter space. Dotted lines indicate regions where the value taken is negative, and solid lines represent positive values. The left panel shows a model with $m_{N_1}=1\,\TeV$ and $|\boldsymbol{h}^\nu_{ij}| = 2.95 \times 10^{-4}$. The right panel displays a model, with $m_{N_1} = 50\:\GeV$ and $|\boldsymbol{h}^\nu_{ij}| = 3.1 \times 10^{-4}$.}
    \label{fig:NeuEvols}
\end{figure}

We now present numerical solutions to the TEs for the TRL model with democratic flavour structure $a=b=c$, and we make use of the {\em dofs} parameterisation given in \cite{Hindmarsh:2005ix}. Additionally, we assume the central value of the sphaleron critical temperature, $T_{\rm sph} = 132 \, \GeV$. Our analysis is limited to heavy neutrino masses above $40\, \GeV$ since, below this scale, the inclusion of thermal corrections to the mass of SM particles leads to phase-space suppression of the decay processes. As a result, the generation of lepton asymmetries must be treated with additional care in regions of the parameter space with very low-scale masses for the heavy neutrinos. Nevertheless, the results presented here will still be valid in order to describe the associated theoretical uncertainties in these regions of the parameter space.

In Figure~\ref{fig:NeuEvols}, we give representative numerical evaluations of the heavy neutrino TEs, with the departure of $\eta^N$ from the equilibrium value $\eta^N_{\rm eq}$ displayed. To carry out these evaluations, we take the masses to be in consecutive resonance with $\mathbb{Z}_6$ symmetric Yukawa structure. In addition, we assume that the heavy neutrino number density is initially in equilibrium, $\Delta(z_0) = \delta(z_0) = 0$, and $\delta\eta^L(z_0) = 0$, where $z_0 = 10^{-2}$. In these two figures, the vertical dotted line indicates the point in the evolution where the sphaleron transitions become exponentially suppressed.
More explicitly, Figure~\ref{fig:NeuEvolsTeV} shows the evolution of the singlet neutrinos when the lightest singlet neutrino is of mass $1\, \TeV$ and the Yukawa coupling scales are $|\boldsymbol{h}^\nu_{ij}| = 2.95 \times 10^{-4}$. As was discussed in Section~\ref{sec:DofsEqu}, at early phases in the evolution, $z < 10^{-1}$, we observe that variations in the {\em dofs} are the dominant phenomena. As a result, the entries of the neutrino matrix take negative values. However, once $z > 10^{-1}$, we see that the decay and inverse decay of the heavy neutrinos push the solution towards the attractor value, and hence the effect on the generated BAU from the {\em dofs} is minimal. 

In Figure~\ref{fig:NeuEvols50}, we now illustrate how the above picture changes for light singlet neutrino masses. For this figure, the lightest singlet neutrino has mass $m_{N_1} = 50\, \GeV$, and the neutrino-Yukawa scale is $|\boldsymbol{h}^\nu_{ij}| = 3.1 \times 10^{-4}$. In this case, the variations in the {\em dofs} have a large impact on the evolution of the different entries of the matrix~$\Delta$,
producing sudden changes that arise from competing effects between the changes in the relativistic {\em dofs} and the changes in the heavy-neutrino number density. Consequently, this non-trivial evolution of $\Delta$ can source significant uncertainties when accurate determinations of the theoretical
parameter space of such low-scale heavy-neutrino models are attempted based on successful leptogenesis.

\begin{figure}[t]
    \centering
    \begin{subfigure}{0.49\linewidth}
        \includegraphics[width=\linewidth]{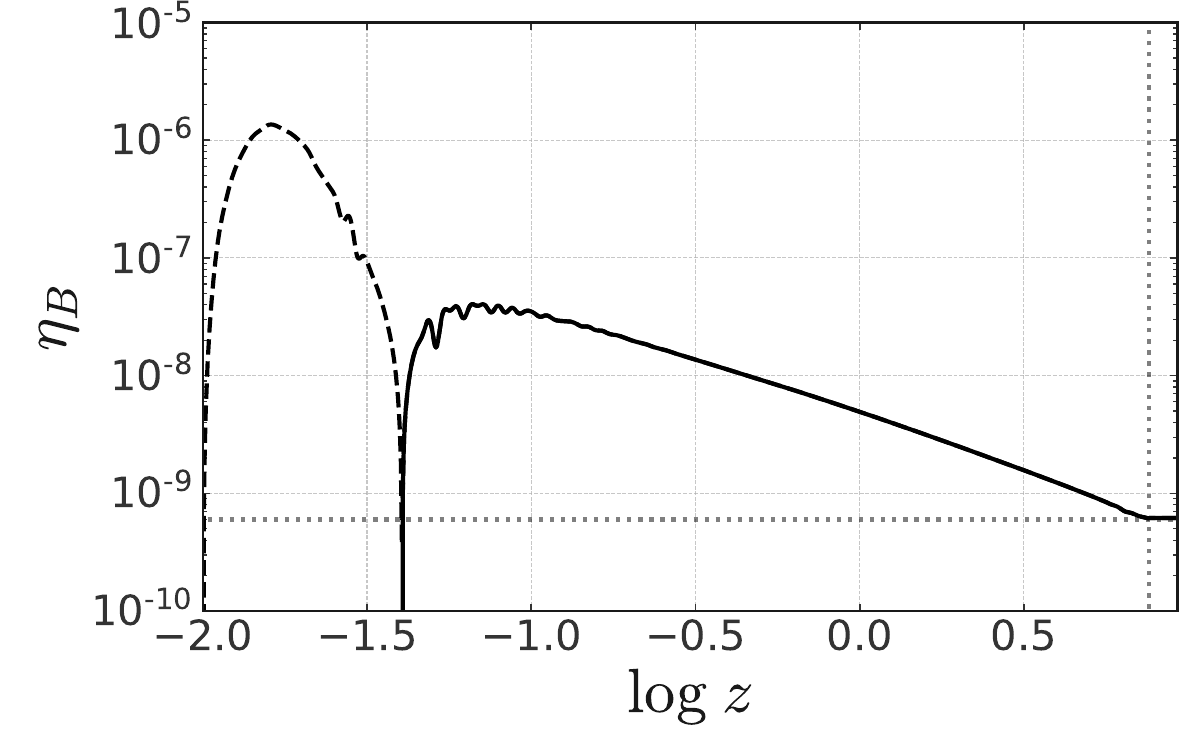}
        \caption{\empty}
        \label{fig:BAUEvolsTeV}
    \end{subfigure}
    \begin{subfigure}{0.49\linewidth}
        \includegraphics[width=\linewidth]{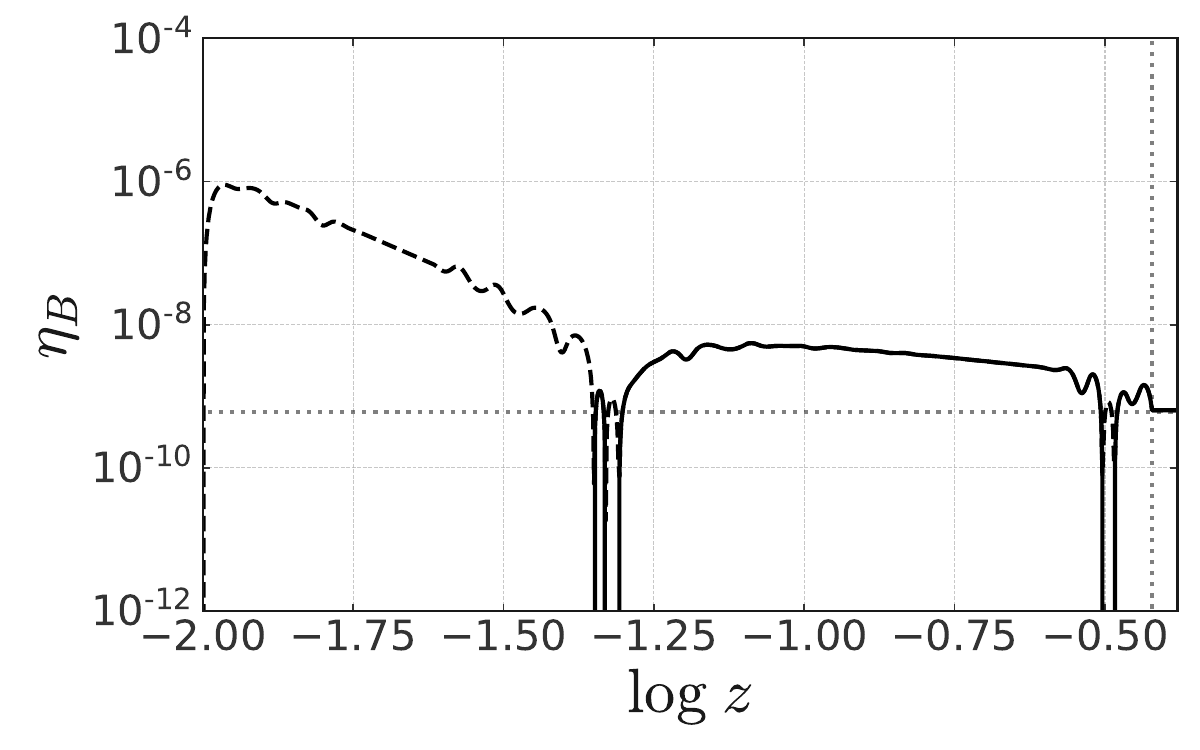}
        \caption{\empty}
        \label{fig:BAUEvols50}
    \end{subfigure}
    \caption{The evolution of the BAU at two points in the parameter space. Dotted lines show where the values are negative, and solid lines indicate positive values. In order to make pertinent comparisons, the left and right panels align with Figures~\ref{fig:NeuEvolsTeV} and~\ref{fig:NeuEvols50}, respectively, with identical initial conditions and model parameters.}
    \label{fig:BAUEvols}
\end{figure}

In Figure~\ref{fig:BAUEvols}, we demonstrate how the above effects feed through into the generated BAU. Again, for these figures, we assume a TRL mass splitting and a $\mathbb{Z}_6$ symmetric Yukawa matrix with a democratic flavour structure. The initial conditions are again chosen assuming that the heavy-neutrino number density is initially in equilibrium, so $\Delta(z_0) = \delta (z_0) = 0$, and $\delta\eta^L(z_0) = 0$ with $z_0 = 10^{-2}$. The vertical dotted line indicates the sphaleron critical temperature and the horizontal line gives the CMB central value of the baryon asymmetry $\eta^B = 6.104\times10^{-10}$. For Figure~\ref{fig:BAUEvolsTeV}, we assume the lightest singlet neutrino has mass $m_{N_1} = 1\, \TeV$ and $|\boldsymbol{h}^\nu_{ij}| = 2.95 \times 10^{-4}$ to align with the inputs of Figure~\ref{fig:NeuEvolsTeV}. We see that in this high-mass regime, the generated BAU is independent of the variations in the {\em dofs} at the start of the evolution, and therefore, the attained values are protected from such phenomena. Moreover, this protection means that any variations within the sphaleron temperature are minimal, as was discussed in Section~\ref{sec:DOfsBAU}. 

Let us now turn our attention to Figure~\ref{fig:BAUEvols50}, where we display the evolution of the BAU when $m_{N_1} = 50\:\GeV$ and $|\boldsymbol{h}^\nu_{ij}| = 3.1 \times 10^{-4}$. As before, this figure should be considered in line with Figure~\ref{fig:NeuEvols50}. In drastic contrast with heavier neutrino models, the generated BAU at this scale is heavily dependent upon the variations in the {\em dofs}. As can be seen from Figure~\ref{fig:BAUEvols50}, there are dramatic variations in the BAU whose value changes sign on multiple occasions. In addition to this less controllable situation, the fact that the sphaleron transitions become suppressed at a significantly lower value of $z$ means that the decays and inverse decays no longer have the time to dominate within the evolution. For such low-scale TRL scenarios, it would require a drastic re-tuning of the mass parameters and CP phases to obtain predictions that lie close to the observed BAU. Hence, we consider that such TRL scenarios are not predictive in standard cosmological settings.

\begin{figure}[t]
    \centering
    \begin{subfigure}{0.49\linewidth}
        \includegraphics[width=\linewidth]{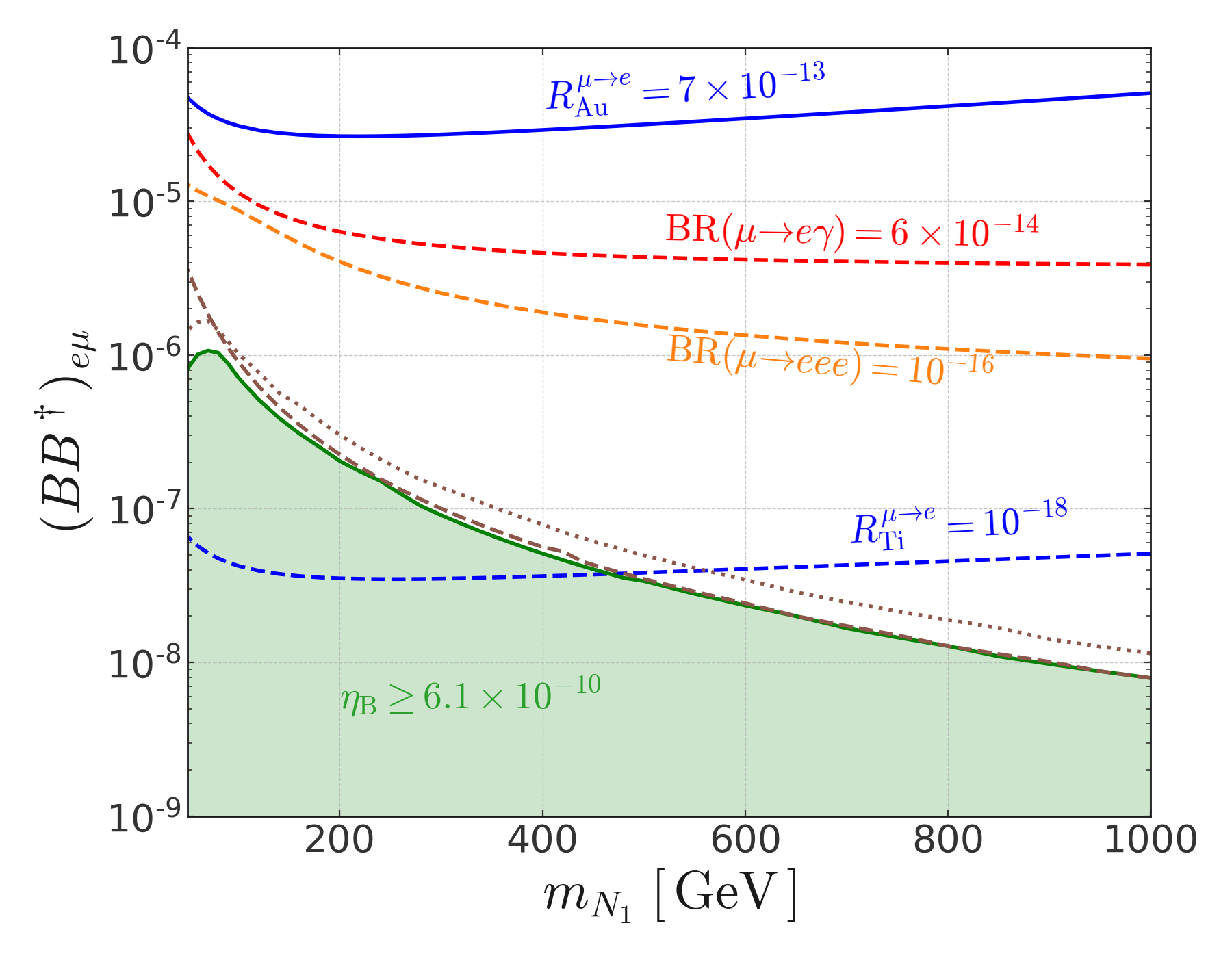}
        \caption{\empty}
        \label{fig:MixCoh}
    \end{subfigure}
    \begin{subfigure}{0.49 \linewidth}
        \hspace*{-3mm}\includegraphics[width=\linewidth]{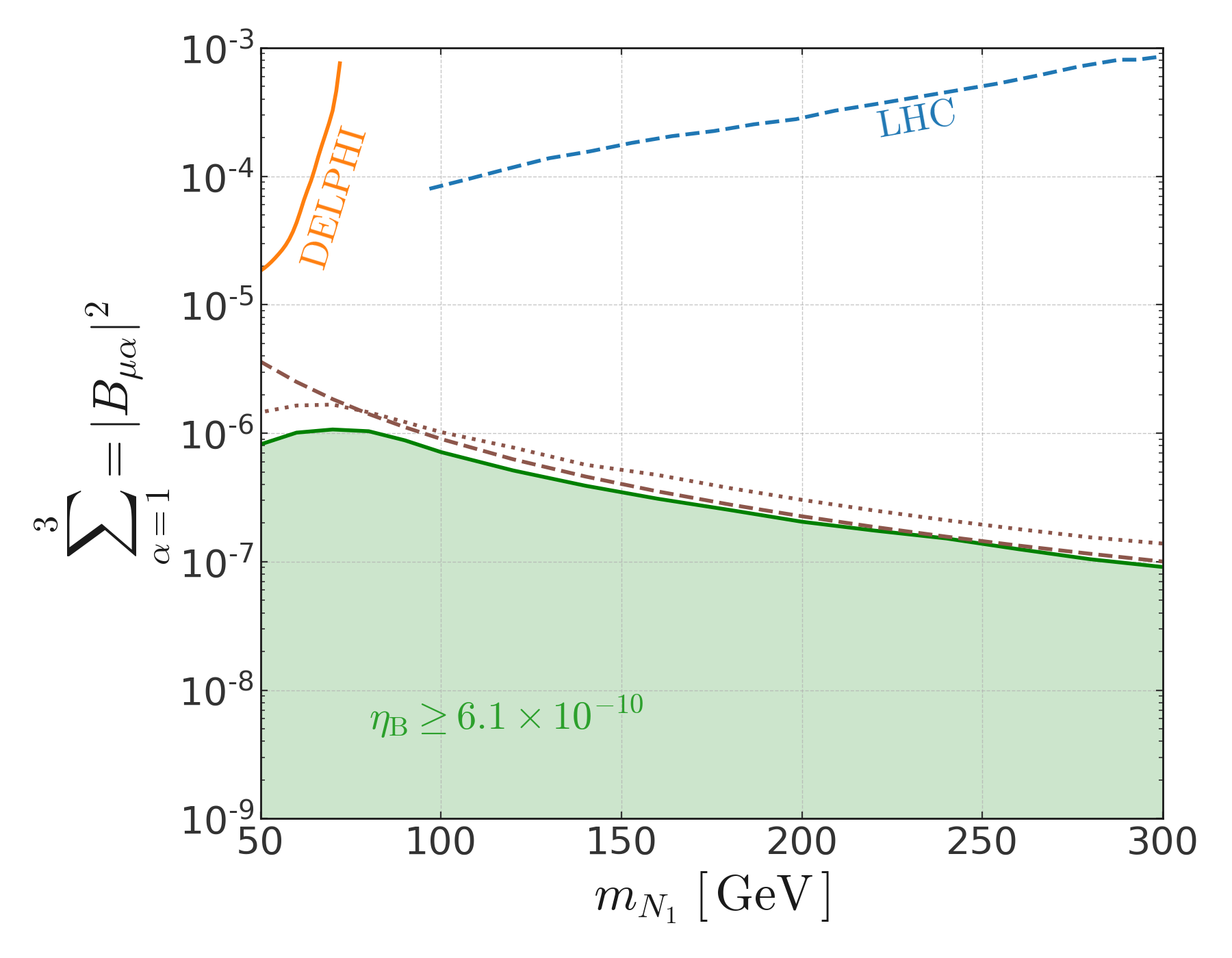}
        \caption{\empty}
        \label{fig:MixColl}
    \end{subfigure}
    \caption{The parameter space for the TRL model with limits from current experiments (solid lines) and projected sensitivity limits (dashed lines) on future experiments. The left panel exhibits the bounds on the parameter space for coherent Lepton Flavour-violating transitions. The red dashed line gives projected bounds on $\mu \to e\gamma$~\cite{MEGII:2018kmf}, the orange dashed line shows $\mu \to eee$ \cite{Blondel:2013ia}, the blue solid line indicates the current bound on cLFV transitions within Gold~\cite{SINDRUMII:2006dvw}, and the blue dashed line gives the projected bound on cLFV transitions within Titanium~\cite{BARLOW201144}. The right panel presents similar bounds for collider experiments. The blue dashed line identifies the projected sensitivity for the LHC running at $\sqrt{s}=14\,\TeV$ ~\cite{Deppisch:2015qwa}. The orange line specifies the current bound on Z-boson decays to singlet neutrinos at DELPHI. Finally, we include brown dashed lines to indicate the parameter space when $\delta_h = 1$, and a brown dotted line which indicates the parameter space in a flavour diagonal TRL model.}
    \label{fig:MixParamSp}
\end{figure}

We now delineate the parameter space for the TRL model and make some pertinent comparisons between this model and special cases of interest. Figure~\ref{fig:MixParamSp} shows the regions of the parameter space which may lead to successful leptogenesis. The solid green line displays the predicted values of the light-to-heavy mixing, $B \simeq \boldsymbol{m}_D \boldsymbol{m}_M^{-1}$ [cf.~(\ref{eq:ChargedCurrentInt})], which should be used to match the observed value for the baryon asymmetry $\eta^{\rm CMB}_B$
given in~\eqref{eq:etaBCMB}. The green-shaded area under the solid line represents regions of the parameter space where the generated BAU exceeds the observed value. In this area of parameter space, the observed baryon asymmetry may still be matched through some additional breaking of the $\mathbb{Z}_6$ symmetry or by a softening of the tri-resonant condition, leading to sub-maximal levels of CP-asymmetry. Alongside the full TRL model, we show the adjusted parameter space when we take the {\em dofs} to be constant and another for a model with no oscillation effects. The former of these is shown in Figure~\ref{fig:MixParamSp} as brown dashed lines and exhibits a clear departure from the TRL model with fully implemented {\em dofs} when in the sub-$100\:\GeV$ regime. For mass scales in excess of $100\:\GeV$, it is apparent that the two lines overlap, in line with the discussion given in Section~\ref{sec:DOfsBAU}. Once the TEs are fully implemented, the variations in the {\em dofs} prevent the generation of baryon asymmetries in the low mass regime. In order to compensate for this, we must reduce the washout of any generated asymmetries, necessitating smaller neutrino-Yukawa couplings. This is then seen in the parameter space as a drop-off in the available mixing. 

In Figure~\ref{fig:MixParamSp} we have also presented the parameter space for the TRL model for which the number density evolution was taken to follow a set of diagonal Boltzmann equations. This is indicated with brown dotted lines in Figure~\ref{fig:MixParamSp}. From this figure, we see that this line follows that of the full model, aside from a global factor, which may be numerically estimated to be around two~\cite{BhupalDev:2014pfm, BhupalDev:2014oar}. It would then appear that the omission of oscillation effects from the evolution seemingly permits additional CP asymmetries whilst maintaining agreement with the observed baryon asymmetry. It also implies that the CP asymmetry associated with oscillation effects interferes destructively with the CP asymmetry associated with singlet neutrino mixing in the decay~\cite{Kartavtsev:2015vto}. Such a result is not entirely unexpected. As has already been discussed, the $\mathbb{Z}_6$ structure of the Yukawa couplings, when combined with a tri-resonant mass hierarchy, provides large CP phases and saturates the available mixing CP-asymmetry. Therefore, we again must reduce the scale of the Yukawa couplings to ease some of the washout effects and generate the necessary baryon asymmetry.

Figure~\ref{fig:MixCoh} shows detection limits on the parameter space from charged Lepton Flavour Violating (cLFV) transitions involving muons. Evidently, the experimental bounds for many experiments still lie far above the parameter space in which we would expect to see cLFV phenomena. However, a small partion of the parameter space may be probed by the PRISM experiment in their future studies of coherent transitions within Titanium. In Figure~\ref{fig:MixColl}, the bounds from collider experiments are displayed. The blue dashed line gives the projected sensitivity bound for the LHC running at $\sqrt{s} = 14\,\TeV$ with integrated luminosity $\mathcal{L}_I = 300\, \rm{fb}^{-1}$, for the process $p\, p \to N \, \ell^\pm j\,j$ \cite{Dev:2013wba, Deppisch:2015qwa}. The orange line gives the estimated sensitivity of the process $Z \to N\nu$ to a $95\%$ confidence level from the DELPHI collaboration \cite{DELPHI:1996qcc}. Moreover, through the decay of $Z$ bosons to charged leptons, it may still be possible to observe some LFV processes~\cite{Abada:2022asx}, although the deviations are expected to be small in TRL models with $\mathbb{Z}_6$ Yukawa structure~\cite{daSilva:2022mrx}. As may be seen in Figure~\ref{fig:MixColl}, many of these experiments do not appear to have the required sensitivity to be able to probe the region of the parameter space which can achieve successful leptogenesis. However, there may be some future experiments, such as the Future Circular Collider (FCC), which can probe the parameter space for masses below $50\:\GeV$~\cite{Blondel:2014bra}.

\section{Conclusions}\label{sec:Concl}
\setcounter{equation}{0}
We have studied how the inclusion of relativistic {\em dofs} influences the predictions for the~BAU in the context of a low-scale Tri-Resonant Leptogenesis model whose neutrino-Yukawa couplings are dictated by a $\mathbb{Z}_6$ symmetry. The TRL model, which we presented in~\cite{daSilva:2022mrx} and reiterated here, is distinctive in that it can produce a vanishing SM neutrino mass spectrum whilst still maintaining large CP phases for the successful generation of the BAU. By considering resummed neutrino-Yukawa couplings, the available CP violation is maximised when the heavy neutrino mass spectrum is taken to be in consecutive resonance, as was detailed in a previous work by the authors in~\cite{daSilva:2022mrx}.

The effect of the relativistic {\em dofs} was analysed by introducing a set of modified Transport Equations, which include additional contributions from variations within the relativistic {\em dofs}. These additional contributions tend to wash out the number density of the heavy neutrinos and drop their number density below its equilibrium value. When this enters the TE for the lepton asymmetry, the generated lepton asymmetry will have the opposite sign to that expected in standard thermal leptogenesis models. At higher scales for which $m_N > 100 \, \GeV$, the variations in the {\em dofs} become a sub-dominant effect, and so thanks to their attractor properties, the TEs can return the generated lepton asymmetry to the expected value in the standard paradigm. 

At low scales, however, where $m_N < 100$~GeV, the $(B+L)$-violating sphaleron transitions, which produce the baryon asymmetry, become exponentially suppressed at a point where the generated lepton asymmetry carries the incorrect sign. This gives rise to the unpleasant feature that a negative BAU gets predicted unless the CP phases in the neutrino Yukawa sector are appropriately re-tuned, according to the sphaleron temperature $T_{\rm sph}$ assumed, the parameterisation of dofs adopted, and the specific value of $m_N$ considered. For regions of parameter space where these re-tunings become excessive, such low-scale TRL models, as well as similar low-scale leptogenesis scenarios, lose their inherent predictive power.

We highlighted that in strong washout regimes, there are always strong attractor trajectories towards the equilibrium. This means that the generated BAU exhibits independence from the initial conditions of the pertinent number densities. Hence, variations in the {\em dofs} play a prominent role regardless of the initial conditions of the heavy-neutrino number density. Since the {\em dofs} phenomena we have studied here contribute an out-of-equilibrium effect, the observed departures from the standard paradigm may be present in many low-scale freeze-out mechanisms, including those to low-scale baryogenesis scenarios and to the generation of a light Dark Matter relic density. However, for weak washout models that rely on the freeze-in framework~\cite{Drewes:2017zyw}, the thermal bath of heavy neutrinos approaches equilibrium slowly since the initial number density vanishes; presumably through some mechanism to be specified that provides low-reheat temperature of a rather convenient scale, not too far above $T_{\rm sph}$. Apart from these strong assumptions on the initial cosmological setting, any deviations in the BAU from the variations in the {\em dofs} are expected to be small.

Along the lines of our work in~\cite{Karamitros:2022oew}, we have extended this previous study to critical-like scenarios within a quasi-thermal bath. In particular${}$, we have demonstrated how, in systems in which the bath is drawn out of equilibrium, the critical phenomena of interest to us may be destroyed. However, in thermo-static scenarios, there still exists the possibility that anomalous behaviours, such as coherence-decoherence oscillations, can be present.

We have delineated in~Figure~\ref{fig:MixParamSp} the parameter space for light-to-heavy neutrino mixings in the TRL model which can successfully reproduce the observed BAU. We see that the inclusion of variations in the relativistic {\em dofs} results in a suppression of the light-to-heavy neutrino mixings for heavy-neutrino masses below $100~\GeV$. Moreover, we find that the heavy-neutrino oscillation effects contribute destructively to the generation of the BAU. Specifically, the diagonal TEs give rise to a bigger parameter space, which is increased by a factor close to two. The parameter space shown in~Figure~\ref{fig:MixParamSp} indicates that many current and future experiments may still be far away from detectable phenomena. However, there still exists some possibility of finding coherent $\mu\to e$ transitions within Titanium at the PRISM experiment. Simple extensions to the TRL model, such as supersymmetry, may offer additional sources of CP violation or additional contributions to the light-to-heavy neutrino mixing. These can bring the predictions for most new-physics observables close to their current experimental limits while achieving successful leptogenesis.

\section*{Acknowledgements}
\noindent
The work of AP and DK is supported in part by the Lancaster-Manchester-Sheffield Consortium for Fundamental Physics, under STFC Research Grants: ST/T$001038$/$1$ and ST/X00077X/$1$. TM acknowledges support from the STFC Doctoral Training Partnership under STFC training grant ST/V$506898$/$1$.

\newpage
\appendix

\renewcommand{\theequation}{\Alph{section}.\arabic{equation}}
\setcounter{equation}{0}
\section{Jordan Forms}\label{App:Jordan}
\setcounter{equation}{0}
In our study of critical scenarios, we are interested in identifying effective Hamiltonians which may reproduce critical phenomena. To this end, we wish to identify the conditions under which the effective Hamiltonian may be brought into a Jordan form~\cite{shafarevich2012linear} with fully degenerate masses and decay widths (or decay rates) \cite{Pilaftsis:1997dr,Karamitros:2022oew}. From the uniqueness of the Jordan Canonical form, we know that for two complex-valued $N\times N$ matrices, $A$ and $B$, if the following criterion is satisfied:
\begin{equation}
  \label{eq:JC}
    \Tr{\lrb{A-B}^k} = 0
\end{equation}
for all $k \in \mathbb{N}$ such that $0<k\leq N$, then $A$ and $B$ have the same complex eigenvalues. However, $A$ and $B$ may not necessarily be related through a similarity transformation. In fact, one has either the trivial case where $A = B$, or if $A\neq B$ and $B$ is diagonal, then $A$ is a Jordan non-diagonalisable $N\times N$ matrix with the same eigenvalues as $B$.

For our purposes, let us consider that the first matrix $A$
is the effective Hamiltonian~$\textrm{H}_{\rm eff}$ for an $N$-level quantum system. Then, $\textrm{H}_{\rm eff}$ can be expanded over the identity, $\mathds{1}_N$ and a set of ${\rm SU}(N)$ generators, $\lrcb{\mathds{1}, \boldsymbol{T}} = \lrcb{\mathds{1}_N\,, T_1\,,\dots \,, T_N}$, with complex coefficients, $H^a \in \mathbb{C}$. In detail, we~have,
\begin{equation}
    \textrm{H}_{\rm eff}\, =\, H^0 \mathds{1}_N\: +\: \boldsymbol{H} \cdot \boldsymbol{T}\;,
\end{equation}
where $\boldsymbol{H} = \big( H^1\,, H^2\,, \dots\,, H^{N^2-1}\big)$ will be called the Hamiltonian vector in the following. Nonetheless, the coefficient of equal interest to us is the zeroth component,
\begin{equation}
    H^0\: =\:  M - \frac{i}{2}\Gamma\;,
\end{equation}
which is the $N$-degenerate complex eigenvalue of $\textrm{H}_{\rm eff}$.  Hence, in order to ensure that all $N$ eigenvalues of $\textrm{H}_{\rm eff}$ are equal to $H^0$, we take the second matrix $B = H^0\mathds{1}_N$ and demand the validity of the $N$-fold criterion~\eqref{eq:JC}, i.e.
\begin{equation}\label{eq:JordCond}
    \Tr{\lrb{\textrm{H}_{\rm eff} - H^0\mathds{1}_N}^k}\, =\, 0\;,
\end{equation}
where $k = 1,\, 2,\, ..., \,N$. From the traceless nature of the $\textrm{SU}(N)$ generators, we see that the condition for $k=1$
is automatically fulfilled. Hence, for $1< k \le N$, 
the sufficient and necessary conditions reduce to
\begin{equation}
    \Tr{\lrb{ \boldsymbol{H} \cdot \boldsymbol{T}}^k}\, =\, 0\;.
\end{equation}
Given an effective Hamiltonian for an $N$-level system, this expression may be used to derive $N$ conditions on the Hamiltonian vector $\boldsymbol{H}$. For $k=2$, we see that this expression gives
\begin{equation}
    \boldsymbol{H}\cdot\boldsymbol{H}\, =\, \sum_{a=1}^N \lrb{H^a}^2\, =\, 0\;.
\end{equation}
Taking the real and imaginary parts of this last relation for an effective Hamiltonian of an $N=2$ level system, the two critical conditions given in~\cite{Karamitros:2022oew} are recovered:
\begin{equation}
   \label{eq:CritQbit}
    \text{(i)}~\Im{\boldsymbol{H}\cdot\boldsymbol{H}} = 0\ \implies\ \boldsymbol{E}\cdot\boldsymbol{\Gamma} = 0\,, \qquad \text{(ii)}~\Re{\boldsymbol{H}\cdot\boldsymbol{H}} = 0\  \implies\ 2|\boldsymbol{E}| = |\boldsymbol{\Gamma}|\,.
\end{equation}
For an $N=3$ level system, one must require, in addition to the two conditions in~\eqref{eq:CritQbit}, the vanishing of the complex-valued sum
\begin{equation}
    \sum_{a,b,c =1}^N d_{abc}\, H^a \, H^b \, H^c \,  =\, 0\;,
\end{equation}
where $d_{abc}$ are the symmetric structure constants of the $\textrm{SU}(N)$ group.

It is now interesting to note a weaker version of the 
criterion stated in~\eqref{eq:JC}. While we permit non-degeneracy of the mass spectrum (with non-negative masses), we still demand that all decay rates be identical to $\Gamma = -2\,\text{Im}\,H^0$~\footnote{The complementary non-critical scenario where all masses are degenerate, but the decay widths different, is quite analogous, and so we will not consider it here.}. From the structure of the effective Hamiltonian, we see that this corresponds to all the imaginary parts of the eigenvalues being equal. Consequently, the weaker condition,
\begin{equation}
    {\rm Tr}\Big[\,{\rm Im} \lrb{A - B}^k\Big]\, =\, 0\;,
\end{equation}
may now be considered in order to determine when $A$ and $B$ will have eigenvalues with identical imaginary parts. Hence, following a similar approach to what was done before but setting now $B = -i\,\Gamma/2\,\mathds{1}_N$, we arrive at the two non-trivial sets of conditions (with $M>0$):
\begin{equation}
  \label{eq:JCweaker}
    \text{(i)}~{\rm Tr}\lrsb{\,{\rm Im} \lrb{\boldsymbol{H}\cdot \boldsymbol{T}}^k}\, =\, 0\,, \qquad \text{(ii)}~\big(NM\big)^k\, \ge\, {\rm Tr} \Big[\,
    \big(M\mathds{1}_N + {\rm Re}\,\boldsymbol{H}\cdot \boldsymbol{T}\big)^k\Big]\, \geq\, N M^k\,,
\end{equation}
where $1 < k \leq N$. Note that the set of all double inequalities in~(ii) of~\eqref{eq:JCweaker} ensures the positivity of all masses. 
In the case that $N=2$, we recover fully the criteria which characterise critical scenarios
\begin{align}
    \text{(i)}~\Im{\boldsymbol{H}\cdot\boldsymbol{H}} = 0\ &\implies\ \boldsymbol{E}\cdot\boldsymbol{\Gamma} = 0\,,\nonumber\\
    \text{(ii)}~~2N(N-1) M^2 \geq\Re{\boldsymbol{H}\cdot\boldsymbol{H}} \geq 0\  &\implies \; 4 M^2 \geq \frac{1}{2}|\boldsymbol{E}|^2 - \frac{1}{8} |\boldsymbol{\Gamma}|^2 \geq 0\, ,
\end{align}
where we have set $k=N= 2$ to obtain the second condition~(ii).
For an $N=3$ level system, we may derive the additional conditions on the geometric structure of the respective effective Hamiltonian,
\begin{align}
   \text{(iii)}&~\sum_{a,b,c =1}^N  \text{Im}\,\Big( d_{abc}\,H^a \, H^b \, H^c \Big)\, =\, 0\;,\nonumber\\ 
   \text{(iv)}&~~N(N^2-1) M^3\, \geq\ \frac{3}{2}\,M\, \Re{\boldsymbol{H}\cdot\boldsymbol{H}}\:
 +\, \sum_{a,b,c =1}^N  \text{Re}\,\Big( d_{abc}\,H^a \, H^b \, H^c \Big)\, \ge\, 0\;,
\end{align}
where $k=3$ was substituted in the last double inequality.

For higher $N$-level systems, extra ladder conditions must be included that become more involved. Nevertheless, they can be straight\-forwardly obtained in terms of products of the antisymmetric and symmetric structure constants, $f_{abc}$ and $d_{abc}$, of the Lie algebra governing the $\textrm{SU}(N)$ group.

\newpage
\bibliography{bibs-refs}{}
\bibliographystyle{JHEP}

\end{document}